\documentclass[%
 reprint,
superscriptaddress,
showkeys,
 amsmath,amssymb,
 aps,
prb,
floatfix,
]{revtex4-2}

\usepackage{environ}
\usepackage{clipboard}
\usepackage{changes}
\usepackage{graphicx}
\usepackage{graphicx}
\usepackage{dcolumn}
\usepackage{bm}
\usepackage{siunitx}
\usepackage{xcolor}
\usepackage{multirow}

\newcommand{\bvec}[1]{\boldsymbol{#1}}

\begin{document}

\title{Negative electronic compressibility in charge islands in twisted bilayer graphene}

\author{Robin J. Dolleman}
 \email{dolleman@physik.rwth-aachen.de}
\affiliation{JARA-FIT and 2nd Institute of Physics, RWTH Aachen University, 52074 Aachen, Germany, EU}
\author{Alexander Rothstein}%
\affiliation{JARA-FIT and 2nd Institute of Physics, RWTH Aachen University, 52074 Aachen, Germany, EU}
\affiliation{Peter Gr\"unberg Institute  (PGI-9), Forschungszentrum J\"ulich, 52425 J\"ulich,~Germany, EU}
\author{Ammon Fischer}
\affiliation{JARA-FIT and Institut f{\"u}r Theorie der Statistischen Physik, RWTH Aachen University, 52074 Aachen, Germany, EU}
\author{Lennart Klebl}
\affiliation{I. Institute of Theoretical Physics, University of Hamburg,
Notkestrasse 9, 22607 Hamburg, Germany, EU}
\author{Lutz Waldecker}
\affiliation{JARA-FIT and 2nd Institute of Physics, RWTH Aachen University, 52074 Aachen, Germany, EU}
\author{Kenji Watanabe}
    \affiliation{Research Center for Functional Materials, 
National Institute for Materials Science, 1-1 Namiki, Tsukuba 305-0044, Japan}
\author{Takashi Taniguchi}
    \affiliation{International Center for Materials Nanoarchitectonics, National Institute for Materials Science,  1-1 Namiki, Tsukuba 305-0044, Japan}%
\author{Dante M. Kennes}
\affiliation{JARA-FIT and Institut f{\"u}r Theorie der Statistischen Physik, RWTH Aachen University, 52074 Aachen, Germany, EU}
\affiliation{Max Planck Institute for the Structure and Dynamics of Matter, Center for Free Electron Laser Science, Hamburg, Germany, EU}
\author{Florian Libisch}
\affiliation{Institute for Theoretical Physics, Vienna University of Technology, 1040 Vienna, Austria, EU}
\author{Bernd Beschoten}
\affiliation{JARA-FIT and 2nd Institute of Physics, RWTH Aachen University, 52074 Aachen, Germany, EU}
\author{Christoph Stampfer}
 \email{stampfer@physik.rwth-aachen.de}
\affiliation{JARA-FIT and 2nd Institute of Physics, RWTH Aachen University, 52074 Aachen, Germany, EU}
\affiliation{Peter Gr\"unberg Institute  (PGI-9), Forschungszentrum J\"ulich, 52425 J\"ulich,~Germany, EU}

\begin{abstract}
We report on the observation of negative electronic compressibility in twisted bilayer graphene for Fermi energies close to insulating states. 
To observe this negative compressibility, we take advantage of naturally occurring twist angle domains that emerge during the fabrication of the samples, leading to the formation of charge islands. 
We accurately measure their capacitance using Coulomb oscillations, from which we infer the compressibility of the electron gas. 
Notably, we not only observe the negative electronic compressibility near correlated insulating states at integer filling, but also prominently near the band insulating state at full filling, located at the edges of both the flat- and remote bands.
Furthermore, the individual twist angle domains yield a well-defined carrier density, enabling us to quantify the strength of electronic interactions and verify the theoretical prediction that the inverse negative capacitance contribution is proportional to the average distance between the charge carriers. 
A detailed analysis of our findings suggests that Wigner crystallization is the most likely explanation for the observed negative electronic compressibility.
\end{abstract}

\maketitle

\section{Introduction}
\begin{figure*}
    \centering
    \includegraphics{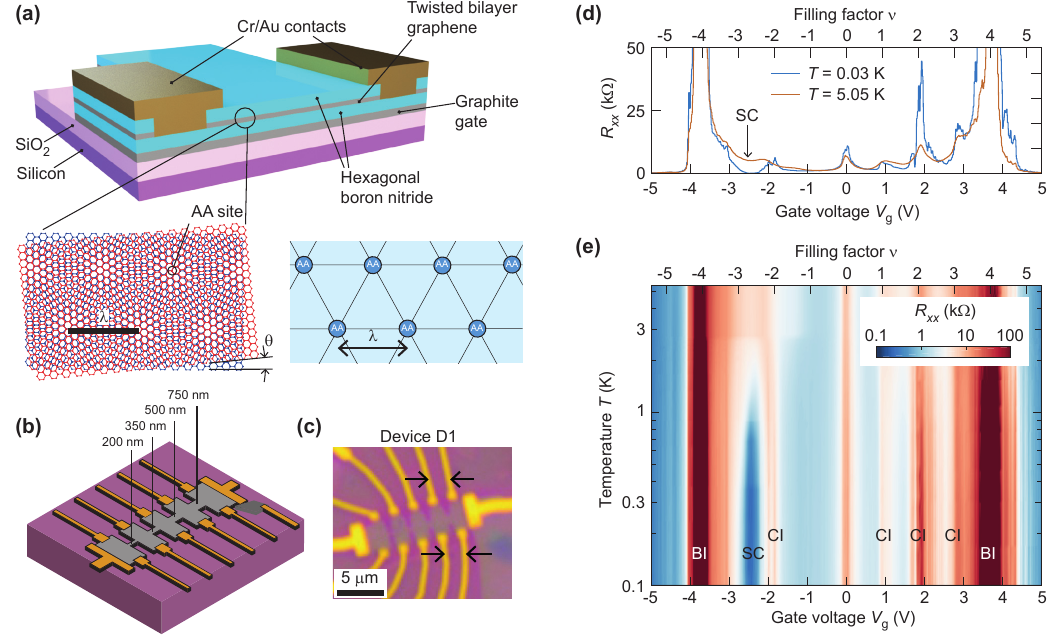}
    \caption{(a)  Schematic cross-section of a tBLG device, including an illustration of the moir\'e lattice forming between the two graphene layers (bottom left) and the resulting stacking order (bottom right). Here, $\theta$ denotes the twist angle between the layers and $\lambda$ the moir\'e wavelength. (b) Illustration of the constricted Hall-bar device D1, with the width of the constrictions labeled. (c) Optical microscopy image of device D1. (d) 4-point resistance as a function of gate voltage $V_g$ and bulk filling factor $\nu$ at two temperatures ($T$), highlighting the fragile superconducting region (SC). (e) 4-point resistance as a function of gate voltage $V_g$ and temperature of the 750-nm-wide constriction, measured with the contacts highlighted by arrows in panel (c). Band insulators (BI), correlated insulators (CI) and a fragile superconducting dome (SC) are visible. }
    \label{fig:D1tempdep}
\end{figure*}
Twisting two layers of graphene creates a moir\'e supercell with enlarged periodicity~\cite{dos2007graphene,morell2010flat,trambly2010localization,bistritzer2011moire}, leading to band insulating states at full filling ($\nu = \pm 4$, where $\nu$ represents the filling factor denoting the number of charge carriers per supercell)~\cite{cao2016superlattice} and flattening of the electronic bands~\cite{morell2010flat,bistritzer2011moire}. 
Near the so-called magic angle of $\approx$~1.1$^{\circ}$, the ratio of Coulomb repulsion to kinetic energy becomes maximized~\cite{kerelsky2019maximized}, giving rise to correlated insulating states at fractional fillings and integer values of $\nu$~\cite{xie2019spectroscopic,cao2018correlated,choi2019electronic,jiang2019charge}.
When the carrier density is slightly tuned away from these integer filling factors, strongly interacting itinerant charge carriers emerge.
The interaction strength of charge carriers is characterized by a dimensionless parameter known as the Wigner-Seitz radius~$r_s$~\cite{falson2022competing,tanatar1989ground}, given by:
\begin{equation}\label{eq:WignerSeitz}
    r_s = \frac{1}{\sqrt{\pi}}\frac{a}{a_B} = \frac{a m^* e^2}{4 \pi^{3/2} \varepsilon_0 \varepsilon_r' \hbar^2}.
\end{equation}
Here, $a = 1/\sqrt{n}$ is the average distance between charge carriers, $n$ the charge carrier density, $a_B$ the (effective) Bohr radius, $\varepsilon_0$ the vacuum permittivity, $\varepsilon_r'$ the effective relative dielectric constant of the material, $\hbar$ the reduced Planck constant, $m^*$ the charge carrier effective mass and $e$ the elementary charge. 
Near an energy gap, the low carrier density (and large $a$) together with a large $m^*$ of itinerant charge carriers results in a large value of $r_s$. 
This high $r_s$ is likely crucial for the formation of strongly correlated electronic phases in twisted bilayer graphene (tBLG), such as the superconducting phases adjacent to the correlated insulating states, the origin of which remains not fully understood~\cite{cao2018correlated,cao2018unconventional,zondiner2020cascade,yankowitz2019tuning,lu2019superconductors,sharpe2019emergent,serlin2020intrinsic}. 
However, precise measurements near these gaps pose challenges due to variations in the twist angle across the sample, which have been identified as a significant source of disorder in tBLG~\cite{tilak2021flat,kerelsky2019maximized,choi2019electronic,zondiner2020cascade,kazmierczak2021strain,schapers2022raman}.
These variations give rise to twist-angle domains within the sample, characterized by relatively uniform twist angles but abrupt transitions at the boundaries~\cite{Lau2022Feb,Uri2020May}.
Since the twist angle dictates the position of moiré-induced energy gaps, their locations vary within the sample, complicating efforts to maintain a uniform itinerant charge carrier density and $r_s$ across the sample geometry.

In this study, we capitalize on the twist angle variations to isolate single twist angle domains, accurately quantify their charging energy, and extract the interaction strength between itinerant charge carriers. 
To achieve this, we utilize tBLG heterostructures with varying sizes and geometries, as described in Section~\ref{sec:samplesetup}. 
We demonstrate that the inherent twist angle variations induce electrostatic confinement near the insulating states of tBLG (Section~\ref{sec:coulomb}). 
These confined regions function as charge islands, and we precisely measure their capacitance using Coulomb oscillations. 
An in-depth analysis of this capacitance reveals a negative electronic compressibility near both the band insulating states in the remote and flat bands, as well as near the correlated insulating states (Section~\ref{sec:negcomp}). 
By fitting a model to the data, we find that the negative capacitance contribution is proportional to the square root of the charge carrier density, consistent with expectations for correlated carriers. 
Furthermore, as the compressibility of the charge island remains unaffected by magnetic fields and is consistent among different bands, we propose that the observed negative compressibility is best explained by the formation of a Wigner crystal (Section~\ref{sec:discussion}). 
 Thus, our technique provides insights into the intriguing properties of itinerant charge carriers in tBLG when the Fermi energy approaches the moiré-induced energy gaps.

\section{Samples and setup}\label{sec:samplesetup}
This study includes a total of six tBLG devices with different twist angles and geometries.
The tBLG is created using either the "tear-and-stack"~\cite{cao2016superlattice} or the "cut-and-stack" technique~\cite{park2021flavour}, generating a moir\'e pattern with periodicity $\lambda$, as illustrated in Fig.~\ref{fig:D1tempdep}(a). 
The tBLG is encapsulated in hexagonal boron nitride (hBN), which serves as an atomically flat protective layer with electrical insulation \cite{Dean2010Oct}. 
To maintain a uniform electric field and minimize electrostatic potential disorder, we utilize a graphite gate, promoting atomically flat interfaces~\cite{Ribeiro-Palau2019Apr,Icking2022Nov}.
For low-resistance one-dimensional contacts, we employ selective reactive ion etching and metallization techniques~\cite{wang2013one,Uwanno2018Aug,Schmitz2017Jun}.
The resulting device structure is depicted in Fig.~\ref{fig:D1tempdep}(a), and an example device is shown in Figs.~\ref{fig:D1tempdep}(b)--\ref{fig:D1tempdep}(c). 
Further details on the fabrication process and devices are available in Appendix~\ref{sec:methods}.

In the main part of this work, we mainly show exemplary results from a specific device labeled D1, which incorporates a Hall-bar structure with constrictions, as shown in Figs.~\ref{fig:D1tempdep}(b)--\ref{fig:D1tempdep}(c).
These constrictions allow us to differentiate between edge and bulk confinement (see Supplemental Material~S2~\cite{Supplemental}).
Within this device, we have implemented constrictions of varying widths: 750 nm, 500 nm, 350 nm, and 200 nm, denoted as C1, C2, C3, and C4, respectively.

To confirm the nature of the 750-nm-wide constriction C1 (Device D1) as tBLG, we analyze the temperature-dependence of the resistance shown in  Figs.~\ref{fig:D1tempdep}(d)--\ref{fig:D1tempdep}(e). 
In this measurement, we can distinctly identify the band insulators (BI) positioned around the filling factor $\nu = \pm 4$. 
By pinpointing their precise locations in gate voltage and utilizing the gate lever arm, we can approximate the twist angle as $\theta \approx 1.02^\circ$ (see Appendix~\ref{sec:methods}).
Moreover, in the vicinity of integer fillings of the partially occupied flat bands, we observe resistance peaks corresponding to fractional fillings $\nu = -2, 1, 2, 3$.
These resistive features align with expectations for correlated insulating (CI) states~\cite{cao2018correlated, stepanov2020untying}.
Close to $\nu = -2.7$, the resistance exhibits a significant reduction with decreasing temperature, although it remains finite.
This behavior indicates the presence of a fragile superconducting state~\cite{park2022robust} in the sample, as evidenced in the Supplemental Material~S1~\cite{Supplemental}.
Furthermore, magnetotransport measurements (Supplemental Material~S1~\cite{Supplemental}) uncover the existence of Chern insulators featuring identical topological invariants to those reported in other studies of tBLG near the magic angle~\cite{nuckolls2020strongly, tomarken2019electronic, stepanov2021competing, saito2021hofstadter, wu2021chern, park2021flavour}.
Consequently, we can conclude that our tBLG sample distinctly exhibits correlated phases akin to those observed in prior investigations near the magic angle~\cite{cao2018correlated, yankowitz2019tuning, lu2019superconductors, sharpe2019emergent, serlin2020intrinsic, stepanov2020untying}.

\section{Charge islands revealed by Coulomb oscillations}\label{sec:coulomb}
\begin{figure}
    \centering    \includegraphics{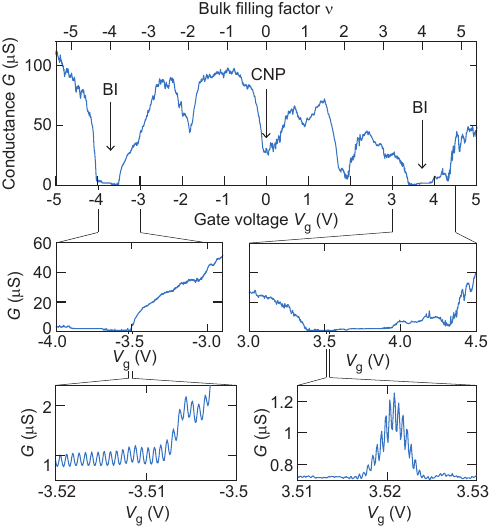}
    \caption{Two-point conductance $G$ as a function of gate voltage $V_g$ over the entire gate voltage range. Zoom-ins close to the band insulating states reveal high-frequency conductance oscillations. }
    \label{fig:observe_oscillations}
\end{figure}
Each experiment which now follows, consists of two-point conductance measurements where we increment the gate voltage in small steps to reveal high-frequency conductance oscillations. 
In Fig.~\ref{fig:observe_oscillations}, we show a conductance ($G$) trace obtained with $0.1$~mV gate voltage steps measured across constriction C1.
The charge neutrality point (CNP) at a filling factor of $\nu = 0$, and band insulators (BI) near full filling at $\nu = \pm 4$ are prominently visible within the trace.
A zoom-in of the traces reveals regular high-frequency conductance oscillations at the transition towards the insulating states of our tBLG samples.
These oscillations are observed in all our tBLG samples that show moir\'e induced energy gaps (see Supplemental Material~S2,~S4,~S5,~S7~\cite{Supplemental}).  

\subsection{Dependence on bias voltage}
\begin{figure}
    \centering
    \includegraphics{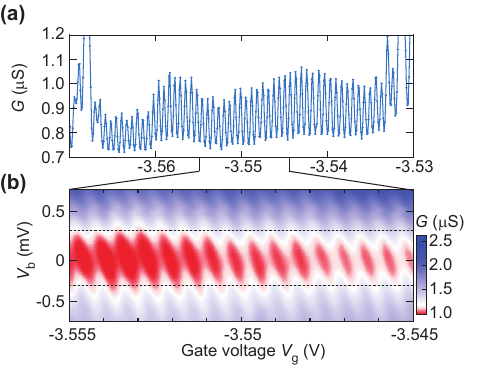}
    \caption{(a) Exemplary 2-point conductance ($G$) trace close to the band insulators or constriction C1. 
    (b) Conductance as a function of gate voltage and bias voltage $V_b$ in a small gate voltage range. 
    }
    \label{fig:biasdependence}
\end{figure}
To identify the origin of the conductance oscillations, we perform gate-dependent bias spectroscopy measurements, as illustrated in Fig.~\ref{fig:biasdependence}.
The combined influence of bias and gate voltage forms a diamond-shaped region where the conductance is suppressed [Fig.~\ref{fig:biasdependence}(b)], a clear manifestation of the Coulomb blockade effect~\cite{stampfer2008tunable}.
In the Coulomb blockade regime, transport is impeded by electrostatic repulsion within a region that confines charge carriers. 
If the bias potential $e V_{\rm b}$ is larger than the charging energy $E_c = e^2/C_{\Sigma}$, where $C_{\Sigma}$ is the total capacitance of the charge island, the Coulomb blockade is lifted~\cite{staring1992coulomb}.
Due to the similarity in magnitude between the AC lockin excitation (100 $\mu$V root mean square) and the step size in bias and gate voltage, the sharp features of the Coulomb diamonds appear blurred. 
Nevertheless, we can extract the bias voltage ($V_{\rm b}$) required to lift the blockade, which is $eV_{\rm b} \approx \pm 0.31$~meV, as indicated by the dashed lines in Fig.~\ref{fig:biasdependence}(b).

The spacing between two Coulomb resonances on the gate-voltage axis is determined by two energy scales: the electrostatic charging energy and the quantum level spacing. 
The remarkable regularity of the measured Coulomb oscillations (Fig.~\ref{fig:observe_oscillations}) suggests that the charging energy dominates over the quantum level spacing. 
Therefore, the distance $\Delta V_{\rm g}$ between the Coulomb resonances is given by $\Delta V_{\rm g} = {e}/{C}$ where $C$ represents the capacitance between the charge island and the graphite back gate~\cite{beenakker1991theory,staring1992coulomb}. 
From Fig.~\ref{fig:biasdependence}, we extract a periodicity of the oscillations $e\Delta V_{\rm g} \approx 0.61$~meV. 
From this analysis, we can conclude that $C \approx C_{\Sigma}/2$, suggesting that half of the total capacitance of the charge island arises from coupling to charge carriers in the source/drain leads.

\subsection{Magnetic field dependence of the oscillations}\label{sec:magfield}
\begin{figure}
    \centering
    \includegraphics{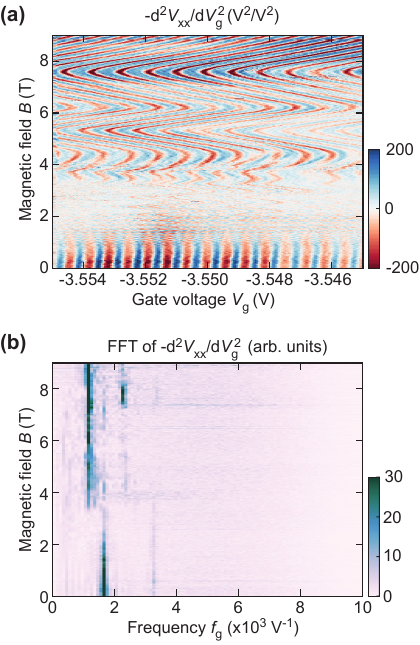}
    \caption{
    Magnetic field dependence of the oscillations. 
    (a) Gradient of the voltage drop (measured in the 4-point configuration) as a function of gate voltage and magnetic field. 
    (b) Power spectral density of the oscillations in panel (a) as a function of magnetic field. 
    }
    \label{fig:Bfield}
\end{figure}
Next, we explore the behavior of the Coulomb oscillations under a perpendicular magnetic field $B$.
We conduct measurements by sweeping the gate voltage in a narrow window while incrementally increasing the magnetic field. 
We observe a significant shift in Coulomb resonance positions [Fig.~\ref{fig:Bfield}(a)], while the spacing between them appears unaffected by the magnetic field. 
A fast Fourier transform (FFT) at each magnetic field reveals two distinct primary frequency components, with higher harmonics arising from the non-sinusoidal shape of the Coulomb resonance [Fig.~\ref{fig:Bfield}(b)]. 
The power spectra's evolution with magnetic field displays a pronounced change in peak amplitude: the frequency component near 1600~V$^{-1}$ decreases, and the component near 1100~V$^{-1}$ becomes more prominent. 
As neither component fully disappears (which is evidenced by the phase coherence of both frequency components in the full magnetic field range, see Appendix~\ref{sec:appBfield}), they must originate from two distinct confined regions within the sample. 
The constant spacing further confirms that the charging energy dominates over the quantum level spacing in the Coulomb blockade effect.
If the quantum level spacing were more significant, the single-particle energies would evolve differently due to the lifting of spin- and valley degeneracy with the magnetic field~\cite{guttinger2009electron,eich2018spin}.

The position of the oscillations in Fig.~\ref{fig:Bfield}(a) exhibits periodicity in $1/B$. 
This behavior, recently been demonstrated in bilayer graphene quantum dots~\cite{banszerus2020electrostatic}, can be explained by a classical electrostatic shift induced by density of states oscillations near the confined region. 
When the Fermi energy matches a Landau level, the density of states in the surrounding area increases, exerting a classical electrostatic force that shifts the energy levels inside the confined regions.
Since an extremum in Fig.~\ref{fig:Bfield}(a) signifies a constant electron density \emph{inside} the confined region, the change in charge density must occur \emph{outside} the confined region.
Therefore, these measurements provide valuable information about the Fermi surface of the carriers in the source/drain leads that couple to the charge island~\cite{banszerus2020electrostatic}, further discussed in Appendix~\ref{sec:appBfield}.

\subsection{Origin of the Coulomb oscillations}\label{sec:originconf}
\begin{figure}
    \centering
    \includegraphics{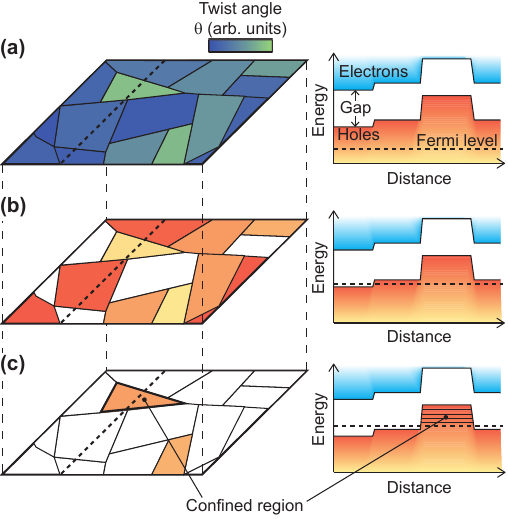}
    \caption{Schematic illustration of twist angle variations and resulting bandstructure variations over the sample geometry.
    (a) Illustration of the formation of twist angle domains over the sample geometry, the local variation of the bandstructure is drawn on the right hand side.
    (b) The Fermi level increases, and in this example the areas with a small twist angle will become insulating since the Fermi level is in their band gap, while other areas remain conducting. 
    (c) Further increasing the Fermi level results in the formation of confinements, where the energy levels are quantized.}
    \label{fig:potentiallandscape}
\end{figure}
The presence of Coulomb oscillations requires an energy gap for charge confinement, such as a band gap, and spatial variations in the position of this gap relative to the Fermi level to create a confining potential for charge carriers. 
In Supplemental Material~S2~\cite{Supplemental}, we investigate the Coulomb oscillations as a function of constriction width on device D1, finding a clear trend of Coulomb oscillations vanishing with decreasing sample width.
This suggests that their origin lies in bulk characteristics rather than edge effects. 
Furthermore, the encapsulation of tBLG with hBN and the use of graphite gating should strongly suppress the effects of charge impurities and other potential disorder~\cite{Icking2022Nov}.
Therefore, we propose that the confinements arise due to variations in the twist angle across the sample geometry~\cite{tilak2021flat,kerelsky2019maximized,choi2019electronic,zondiner2020cascade,kazmierczak2021strain,schapers2022raman}, leading to the formation of twist angle domains over the sample~\cite{Lau2022Feb,Uri2020May}, as illustrated in Fig.~\ref{fig:potentiallandscape}(a).
 Local variations in the twist angle result in local changes in the energy gap offset relative to the Fermi level [see the potential landscape on the right-hand side of Fig.~\ref{fig:potentiallandscape}].
 Therefore, as the Fermi level approaches the band gap, certain domains become insulating earlier than others [Fig.~\ref{fig:potentiallandscape}(b)].
 Close to the insulating state, individual domains may remain conducting while the surrounding area has already become insulating [Fig.~\ref{fig:potentiallandscape}(c)].
 This scenario leads to individual charge islands and the observation of conductance oscillations as a function of gate voltage.

 However, this leaves the question of how these charge islands are accessible in the transport experiment.
An in-depth analysis of the quantum oscillation frequency in Appendix~\ref{sec:appBfield} reveals that the carriers tunneling in and out of the confinement exhibit a complete absence of moiré-induced energy gaps. 
This observation points toward an important role of the boundaries between twist angle domains.
Within these boundaries, strong disorder may be present on the length scale of the moiré superlattice. 
Consequently, locally, the moiré-induced gap may vanish, leaving charge carriers which tunnel in and out of the confinement.
Since these boundaries are expected to appear over the entire sample geometry, we propose that they form a network that allows to probe the local charge confinement in electronic transport experiments. 

\section{Negative electronic compressibility}\label{sec:negcomp}
\begin{figure*}
    \centering
    \includegraphics{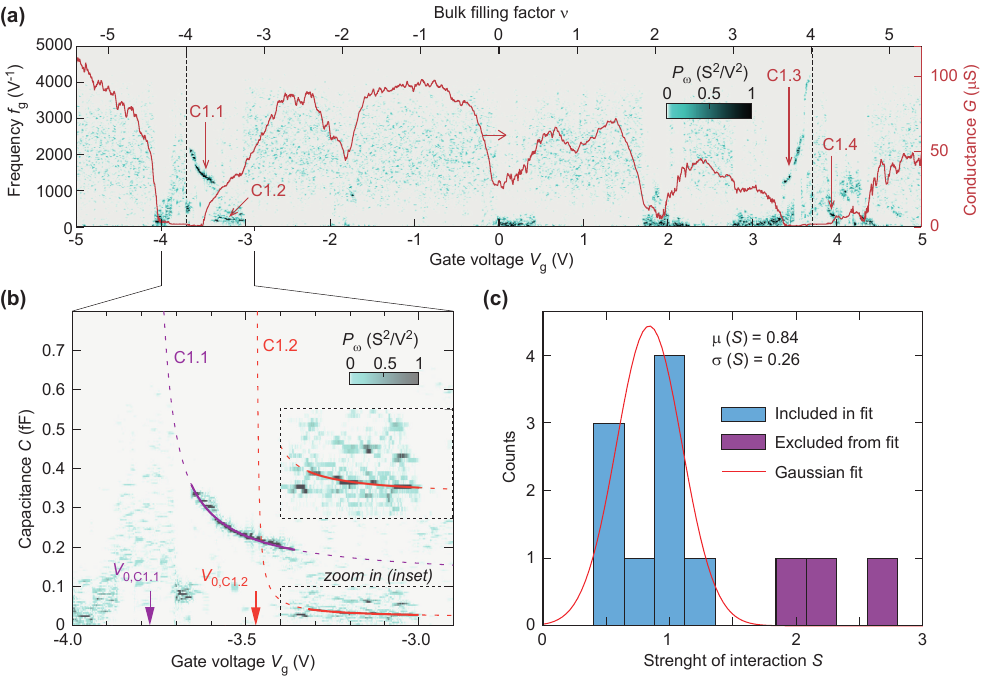}
    \caption{Observation of negative compressibility and determining the interaction strength $S$. (a) Power spectrum $P_{\omega}$ versus capacitance $C$ and gate voltage $V_g$ on constriction C1. The conductance trace (right vertical axis) is included for easy comparison. (b) Zoom-in of $P_{\omega}$ near the band insulators at $\nu = - 4$. Equation~\eqref{eq:slopeequation} is fit to two frequency components. The inset is scaled vertically by a factor 5 to highlight component C1.2. (c) Histogram of all the values of $S$ determined from fitting and determining the mean of $S$. }
    \label{fig:frequencytuning}
\end{figure*}
Next, we study the spacing of the Coulomb oscillations $\Delta V_g$ to determine the capacitance of the charge islands. 
Since $\Delta V_{\rm g} = e/C$, the oscillation frequency $f_{\rm g}$ can be expressed as $f_{\rm g} = {C}/{e}$, directly proportional to the back gate capacitance of the charge island. 
To facilitate the analysis, we convert conductance traces to the frequency domain by calculating power spectra $P_{\omega}  = |\mathcal{F} \{ \mathrm{d} G/ \mathrm{d} V_g \} |^2$, where $\mathcal{F}{}$ represents the Fourier transform (for more details, see Appendix~\ref{sec:methods}).
The resulting gate-voltage-dependent power spectra, presented in Fig.~\ref{fig:frequencytuning}(a) for device C1, reveal multiple distinct frequency components near the insulating states (labeled as C1.1 to C1.4), each corresponding to a single charge island.
While the oscillation period remains regular within small gate voltage ranges ($< 10$~mV), a continuous frequency tuning is evident on larger voltage scales.
Across all tBLG devices in this study, we consistently observe an increase in frequency (and capacitance) as the Fermi level approaches the moiré-induced energy gap, representing a central result of our work (additional examples are presented in Supplemental Material~S2,~S4,~S5,~S7~\cite{Supplemental}).

To elucidate the observed change in capacitance, we first consider the geometric capacitance $C_g$ of the charge island within a parallel-plate capacitor model, given by $C_g = {\varepsilon_0 \varepsilon_r A}/{d}$, where $A$ is the area of the island, and $d$ is the distance between the tBLG and the graphite gate.
Based on the twist angle domain model outlined in Section~\ref{sec:originconf} and Fig.~\ref{fig:potentiallandscape}, we anticipate that $A$ and $C_g$ should remain constant as a function of gate voltage if the charge island consists of a single domain.
However, if the charge island comprises multiple domains, we expect to observe a step-wise reduction of $A$ and the capacitance.
Alternatively, if the confinement arises from other types of potential disorder \cite{gold2020scanning}, we anticipate a continuous reduction of the area $A$, as more of the surrounding area should become insulating when the Fermi level enters the bandgap in the surrounding bulk. 
Given that these scenarios are inconsistent with our observations, we can conclude that the observed increase in capacitance is indicative of a negative compressibility of the charge carriers. 

\subsection{Models for the negative capacitance contribution}
To investigate the negative compressibility contributing to the observed increase in capacitance, we consider two scenarios where negative compressibility can arise.
First, we consider the exchange interaction in an electron gas, where the reduced likelihood of finding charge carriers with the same spin at the same position creates an "exchange hole" due to the opposite background charge~\cite{Eisenstein1992Feb,tanatar1989ground,steffen}. 
The resulting negative interaction energy $E_{i,X}$ is given by:
\begin{equation}\label{eq:exchange}
E_{i,X} = - \frac{4}{3} \frac{e^2 A  n^{3/2}}{\sqrt{2 \pi} \varepsilon_0 \varepsilon_r'} \left[ (1+\xi)^{3/2} + (1 - \xi)^{3/2} \right],
\end{equation}
where $\xi$ represents the polarization of magnetic moments, ranging from 0 (unpolarized) to 1 (fully polarized).
Secondly, we consider the case of a Wigner crystal, where charge carriers minimize their potential energy by forming a solid phase with a triangular lattice~\cite{meissner1976stability}.
The resulting negative interaction energy $E_{i,W}$ is given by~\cite{meissner1976stability,bonsall1977some,bello1981density,ruzin1992pinning,fu2015correlation,joy2021wigner}:
\begin{equation}\label{eq:interactionenergy}
E_{i,W} = -\frac{\eta_T e^2 A n^{3/2}}{8 \pi \varepsilon_0 \varepsilon_r'},
\end{equation}
where $\eta_T = 3.92$ is a numerical constant associated with the triangular lattice.

In both scenarios, the negative interaction energy $E_i$ leads to a negative (thermodynamic) electronic compressibility $\kappa$, expressed as $\kappa^{-1} = \mathrm{d}\mu/\mathrm{d}n = (1/A)(\mathrm{d}^2 E_i/\mathrm{d} n^2)$, where $\mu$ is the electrochemical potential.
The negative compressibility contributes to a negative capacitance contribution $C_i$, where $C_i^{-1} = \kappa^{-1}/(Ae^2)$, increasing the total capacitance $C$ beyond the geometric contribution $C_g$ according to the relation~\cite{skinner2010anomalously,fu2015correlation,wang2013negative,ilani2006measurement}:
\begin{equation}\label{eq:capacitancerelation}
C^{-1} = C_g^{-1} + C_i^{-1} =C_g^{-1} + \frac{1}{A^2e^2}\frac{\mathrm{d}^2 E_i}{\mathrm{d}n^2}.
\end{equation}
Since the carrier density dependence of both interaction energies is the same [Eqs.~\eqref{eq:exchange}--\eqref{eq:interactionenergy}], we can use Eq.~\ref{eq:capacitancerelation} to obtain a general model for the capacitance, including the effect of correlated itinerant charge carriers:
\begin{equation}\label{eq:slopeequation}
        \frac{1}{C} = \frac{d}{\varepsilon_0 \varepsilon_r A} - \frac{S}{ \varepsilon_0 \varepsilon_r' A \sqrt{|n|}},
    \end{equation}
where the dimensionless parameter $S$ characterizes the strength of the correlation, which can be experimentally determined through fitting.
The expressions and values for $S$ for the different theoretical scenarios can be found in Table~\ref{tab:Sparam}.
To fit Eq.~\ref{eq:slopeequation} to our data, we express the carrier density as $|n| = \alpha |V_g - V_0|$, where $V_g$ is the gate voltage, $V_0$ is the gate voltage where the itinerant charge carrier density is zero, and $\alpha$ is the lever arm. 
We determine $\alpha$ from the Landau levels that emerge in magnetotransport (see Supplemental Material~S1~\cite{Supplemental}). 

\begin{table}[]
\caption{Expressions for the $S$ in different correlated electron models compared to the experimental value. The abbreviation WC denotes the Wigner crystal scenario, while HF denotes the Hartree-Fock model in the exchange scenario. \label{tab:Sparam} }
\centering
\begin{ruledtabular}
\begin{tabular}{ccc } 
     & $S $ & Value \\[5pt] \hline
\multirow{2}{*}{WC} &
  \multirow{2}{*}{$\displaystyle \frac{3\eta_T}{32\pi }$}  &
  \multirow{2}{*}{$\num{0.12}$} \\
           &   &                             \\[5pt] 
\multirow{2}{*}{HF, unpolarized} &
  \multirow{2}{*}{$\displaystyle \frac{2}{\sqrt{2 \pi} }$} &
  \multirow{2}{*}{$\num{0.80}$} \\
           &   &                             \\[5pt] 
\multirow{2}{*}{HF, polarized} &
  \multirow{2}{*}{$ \displaystyle \frac{2.828}{\sqrt{2 \pi}}$} &
  \multirow{2}{*}{$\num{1.13}$} \\
           &   &                             \\[5pt] 
Experiment & - & $\num{0.84} \pm \num{0.26}$
\end{tabular}
\end{ruledtabular}
\end{table}

\subsection{Determining the strength of the interaction}
Figure~\ref{fig:frequencytuning}(b) shows the power spectrum of constriction C1 close to the band gap at $\nu = -4$. 
We identify and determine the maxima of two frequency components and fit Eq.~\eqref{eq:slopeequation} to extract $S$. 
The resulting excellent fit demonstrates that the $n^{-1/2}$ dependence in our model provides a good description of the gate-voltage-dependent frequency observed in the experiments.
We perform this analysis on 12 frequency components from five different samples that are close to the magic angle: constrictions C1, C2 and C3 on Device D1; and Devices D2 and D3. 
The obtained values of $S$ are summarized in Fig.~\ref{fig:frequencytuning}(c) and the complete fitting results are presented in the Supplemental Material~ S3~\cite{Supplemental}.
The mean value of $S$ is found to be $\mu(S) = 0.84$ with a standard deviation of $\sigma(S) = 0.26$. 
Three outliers in Fig.~\ref{fig:frequencytuning}(c) were excluded from calculating the mean value due to relatively large uncertainty bounds resulting from the fitting procedure.

\subsection{Analysis with a fixed strength of the interaction}
\begin{figure*}
    \centering
    \includegraphics{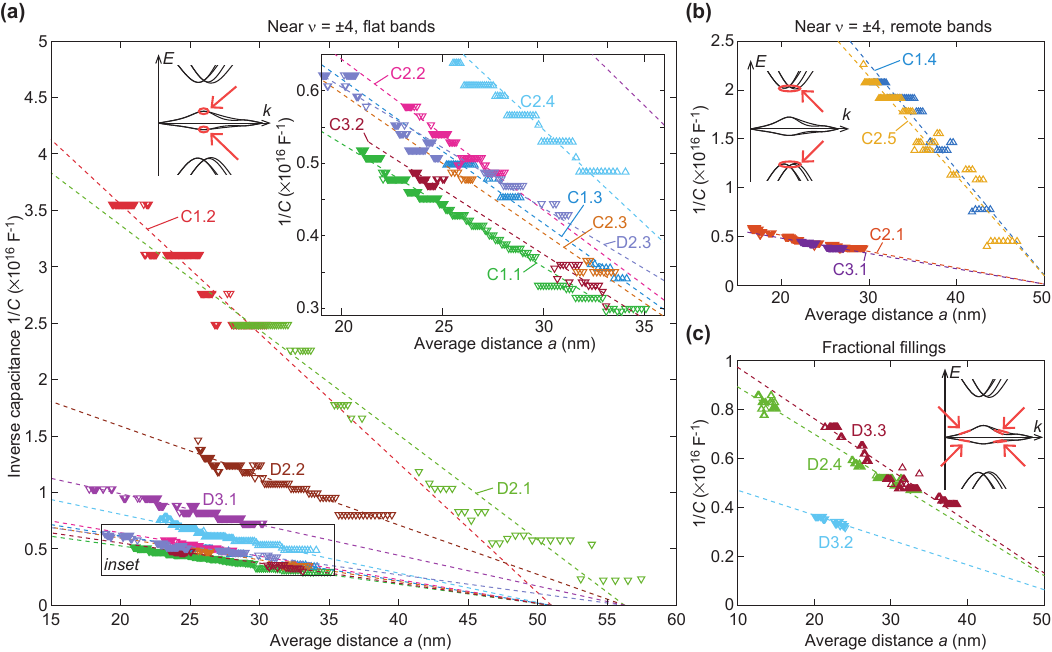}
    \caption{Maxima in the power spectrum used to fit Eq.~\eqref{eq:slopeequation}, plot with $1/C$ on the vertical axes and the average distance between carriers $1/\sqrt{n} = a$ on the horizontal axes such that Eq.~\eqref{eq:slopeequation} becomes a straight line. The fits to each frequency component are made with a fixed value of $S = 0.84$. The frequency components are divided into panels such that: (a) shows components found near $\nu = \pm 4$, on the side of the flat band (see the simplified bandstructure in the inset) (b) shows components found near $\nu = \pm 4$, on the side of the remote band and (c) shows frequency components which are found near fractional fillings of the superlattice.   }
    \label{fig:fits}
\end{figure*}
Next, we refine our model by fixing the parameter $S$ to the mean value of $S = 0.84$, i.e., we account for all our data from devices close to the magic angle (the constrictions C1, C2, C3 and devices D2, D3) and band gap transitions using one single, fixed value of $S$. 
This allows us to fit the subset of frequency components that occur within a narrow gate voltage range, expanding our dataset to include 18 frequency components.
Note that in this refined model, both the vertical offset and steepness of the curve are solely determined by the size of the confinement $A$, while $V_0$ determines the horizontal position.
The fitting results are presented in Figure~\ref{fig:fits}, where the vertical axis represents the inverse capacitance $1/C$, and the horizontal axis represents the average distance between the charge carriers $a = \alpha \sqrt{|V-V_0|}$.
Each frequency component is labeled using the format "xx.y", where "xx" is the constriction or device number, and "y" is the number of the frequency component.
According to Eq.~\eqref{eq:slopeequation}, the resulting maxima should fall along the straight line determined by the fit.
Remarkably, using a fixed value of $S$ provides excellent fits to all the data, regardless of the filling factor or the band where the frequency component is observed.
The remaining plots of the frequency components, including the fits and the fitted values of $A$ and $V_0$, are presented in the Supplemental Material~S3~\cite{Supplemental}.

It is noteworthy that we also find periodic Coulomb oscillations that are well-described by Eq.~\eqref{eq:slopeequation} close to the insulating states that form at fractional superlattice fillings [Fig.~\ref{fig:fits}(c)]. 
These include components D2.4 found near $\nu = 2$, D3.2 near $\nu = -2$ and D3.3 near $\nu = 3$.
The latter findings clearly demonstrate that a single-particle band gap is not necessary for confined regions to arise in the sample; instead, the charge gap arising from the correlated insulating states is sufficient. 
The fact that they are well-described by Eq.~\eqref{eq:slopeequation}, assuming a zero itinerant charge carrier density at $V_0$ in the partially filled band, is also in-line with the opening of a correlation-induced energy gap near the integer fillings of the superlattice.

\subsection{Size and twist angle of the charge islands}
Within this subsection, we delve deeper into the results of our modeling and fitting analysis to evaluate their alignment with the scenario we proposed in Section~\ref{sec:coulomb} involving twist angle domains.
By comparing $V_0$ to the voltage where we pinpoint the center of the band gap in the bulk, we can estimate the deviation of the twist angle.
We do this for constriction C1, and find that components C1.1, C1.2, C1.3, C1.4 deviate $-0.023^{\circ}$, $0.016^{\circ}$, $0.011^{\circ}$ and $-0.003^{\circ}$ respectively from the bulk value of 1.022$^{\circ}$. 
These twist angle variations are in-line with observations in the literature~\cite{Uri2020May}. 
This shows that minute twist angle variations of only a few percent from the average bulk value suffice to produce the confinements observed in this work. 

\begin{figure}
    \centering
    \includegraphics{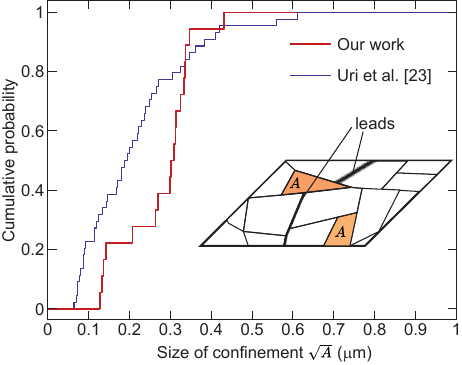}
    \caption{Empirical cumulative distribution function (CDF) of the size of the confinements $\sqrt{A}$ compared to the sizes of twist-angle domains estimated from the results of \citeauthor{Uri2020May}~\cite{Uri2020May}. Insert shows a schematic of the confinements and the "leads" along the twist angle boundaries where transport is thought to occur. }
    \label{fig:size_distribution}
\end{figure}
Figure~\ref{fig:size_distribution} displays the empirical cumulative distribution function (CDF) of the sizes ($\sqrt{A}$) of the 18 charge islands extracted from our fits on all frequency components. 
These sizes are consistent with the dimensions of our samples and similar to the experimental twist angle maps obtained by \citeauthor{Uri2020May}~\cite{Uri2020May}. 
A notable difference is that we find no substantially smaller ($\sqrt{A} < 125$ nm) or larger ($\sqrt{A}>425$ nm) areas in our experiments. 
We note that small areas are usually embedded into a larger twist angle domain that exhibits more uniformity, causing twist angle boundaries that can couple to the charge island to be absent \cite{Uri2020May}.
On the other hand, large domains are less likely to form a detectable charge island, since the large circumference makes it less probable that the entire surrounding region is insulating at the same time.
Considering these factors, our size distribution aligns reasonably well with the literature, validating our model and the obtained value of  $S \approx 0.84$.

\section{Nature of the correlated state}\label{sec:discussion}
Our work provides compelling evidence for a negative capacitance contribution within confined regions of tBLG.
However, the nature of the correlated state remains uncertain. 
The analysis has established the $1/\sqrt{|n|}$ dependence of the inverse capacitance, consistent with both an exchange contribution or a Wigner crystal phase. 
Despite the experimental value of $S \approx 0.84$ (Table~\ref{tab:Sparam}) closely resembling that associated with exchange contributions in an unpolarized electron gas ($S = 0.80$), several experimental observations raise doubts:

(1) The observation of regular Coulomb oscillations suggests that the spin- or valley degree of freedom is not significant for the addition energy in the confined region.

(2) Substantial changes in the polarization of magnetic moments with magnetic field, including the disappearance of correlated insulating phases at high magnetic fields (see Supplemental Material~S1~\cite{Supplemental}), indicate that the Zeeman energy exceeds the exchange-induced energy gap. At a magnetic field of 9 T, the Zeeman energy in the order of 1 meV should dominate over other spin contributions, leading to a fully polarized electron gas. Consequently, if the electron gas is not polarized at zero field, the change in polarization between 0~T and 9~T would lead to a frequency increase. For example, we would expect a frequency increase of 72\% as $\xi \rightarrow 1$ for the 1100~V$^{-1}$ component in Fig.~\ref{fig:Bfield}(b). Instead, an indication of a change in the Coulomb oscillation frequency is observed in none of our measurements, additional examples are included in the Supplemental Material~S9~\cite{Supplemental}.

(3) If the charge carriers are fully out-of-plane polarized at $B = 0$~T, no change in the Coulomb oscillation frequency is anticipated in the exchange scenario. 
However, no manifestation of such ferromagnetism such as an anomalous Hall effect~\cite{Nagaosa2010May} is observed, nor do we observe hysteresis when sweeping the magnetic field from -9 to 9~T and vice versa (see Supplemental Material~S8~\cite{Supplemental}).

(4) The carrier density dependence of the polarization of magnetic moments in the fluid phase is expected to be strong~\cite{zondiner2020cascade}, potentially leading to significant deviations from the observed $1/\sqrt{n}$-dependence of the inverse capacitance in the exchange scenario. 

(5) Describing frequency components with the same value of $S$ regardless of whether they are found near a correlated or band insulator is unexpected in the case of exchange contributions, according to theoretical simulations~\cite{Rai2023Sep}, because the spin/valley polarization varies for each energy gap that opens in tBLG \cite{zondiner2020cascade}.

(6) Theoretical models that take short-range interactions into account indicate that strong negative contributions to the capacitance are not anticipated close to the band insulating states at $\nu = \pm 4$~\cite{Rai2023Sep}.

The absence of clear signatures expected in the exchange scenario and the consistent value of $S$ across different bands suggests that a Wigner crystal phase may provide a better explanation for the observed negative compressibility. 
The exchange energy in a Wigner crystal follows a scaling law of the form $E_{\rm X} \propto \exp{-\gamma \sqrt{r_s}}$~\cite{spivak2004phases,roger1984multiple}, where $\gamma$ is of the order 1.
As a consequence, the interaction energy of a Wigner crystal does not rely on the polarization $\xi$, providing an explanation for the six observations listed above. 
The higher value of $S$ than expected for a Wigner crystal may be related to the average distance $a$ being in the same order to the moir\'e wavelength $\lambda$, causing a fraction of charge carriers to obtain even lower energy states within the non-uniform potential landscape provided by the moir\'e lattice. 
Importantly, as shown in the Appendix~\ref{sec:appbeta}, this effect also leads to the same $1/C \propto 1/\sqrt{n}$ dependence, consistent with the experimental findings.
Further investigations, considering additional factors and refinements to the model, may provide a more detailed understanding of the presently observed correlated state in tBLG.

\section{Conclusions}
This study demonstrates a negative electronic compressibility in confined regions of tBLG.
The observed dependence of the inverse capacitance on the charge carrier density, characterized by $1/C \propto 1/\sqrt{n}$, aligns with the presence of strongly correlated itinerant charge carriers. 
The magnetic field dependence and the similarity of the correlation strength parameter $S$ across different bands suggest a limited role of exchange contributions, pointing towards Wigner crystallization as a likely explanation for the observed negative compressibility.
These findings highlight the role of naturally occurring electrostatic confinements in tBLG, which enable precise investigation of the ground state energy of correlated states in this material.

Our research reveals two distinct phenomena rooted in the moiré physics of tBLG. 
The first phenomenon involves electrostatic confinements arising from variations in the twist angle. 
This effect hinges solely on the presence of moiré-induced energy gaps, while the flatness of the band is not important. 
Therefore, this phenomenon is not exclusive to samples near the magic angle, as evidenced in tBLG samples far from this angle (Supplemental Material~S4,~S6~\cite{Supplemental}).
The second phenomenon is the manifestation of negative compressibility within these confinements. 
In this case, the band flattening in tBLG plays a crucial role, since $r_s$ is required to be large.  
This gains support from control experiments, particularly one conducted on a sample where the twisted bilayer has relaxed. 
In this case, we observe the absence of moiré minigaps, while the charging spectrum of a confinement in this sample shows no negative compressibility (Supplemental Material~S6~\cite{Supplemental}).
Furthermore, we extend our investigation to samples with twist angles significantly deviating from the magic angle ($\theta \sim 0.65^{\circ}$ in Supplemental Material~S5 and $\theta > 1.29^{\circ}$ in Supplemental Material~S7~\cite{Supplemental}).
While these samples do exhibit an increase in frequency as the Fermi level approaches an energy gap, their evolution with gate voltage deviates from the trend specified in Eq.~\eqref{eq:slopeequation}.
In these instances, the parameter $r_s$ may not be sufficiently large to induce negative compressibility in the full gate voltage range where the confinement is formed, a factor that can be attributed to the reduced band flattening compared to the magic angle.

While it is commonly believed that correlated effects are not significant in the remote bands due to the absence of correlated insulating states~\cite{Das2021Jun}, our observation of negative compressibility in these bands [Fig.~\ref{fig:fits}(b)] challenges this notion. 
An analysis of the band structure, including a Hartree-Fock correction, suggests that the observed negative compressibility may be attributed to a significant correlation-induced flattening of the band when the filling factor exceeds $|\nu| = 4$ (see Appendix~\ref{sec:bandstructure}).

The size of a typical charge island, as illustrated in Fig.~\ref{fig:size_distribution}, is notably larger than the moiré wavelength ($\lambda = 13.8$ nm for a 1.02$^{\circ}$ twist angle). 
This suggests that the periodicity of the moir\'e potential remains a crucial factor within the charge island, and the negative compressibility is not solely a consequence of electrostatic confinement.
Our experimental evidence strongly supports this idea, revealing a significant suppression of the kinetic energies of carriers within the confinement compared to single-layer or Bernal-stacked bilayer graphene. 
In essence, we have demonstrated how moir\'e-induced confinement can offer quantitative insights into the interaction energies of correlated effects in tBLG by accurately measuring the local charging energy. 
This is the principal achievement of our work and has significant implications for understanding and manipulating correlated phases in moiré materials.

~\\
~\\
The source data and Matlab code underlying this paper are available at Ref.~\cite{zenodo} upon publication of this manuscript.

\begin{acknowledgements}
The authors thank C. Volk and J. Sonntag for experimental support; F. Volmer, T. Ouaj and M. Schmitz for assistance during sample fabrication; S. Trellenkamp and F. Lentz for electron-beam lithography; and L. Gaudreau and M. Morgenstern for discussions. 
This work was supported by the FLAG-ERA grants 436607160 2D-NEMS, 437214324 TATTOOS and 471733165 PhotoTBG by the Deutsche Forschungsgemeinschaft (DFG, German Research Foundation) and by the Deutsche Forschungsgemeinschaft (DFG, German Research Foundation) under Germany's Excellence Strategy – Cluster of Excellence Matter and Light for Quantum Computing (ML4Q) EXC 2004/1 – 390534769, by the Deutsche Forschungsgemeinschaft (DFG, German Research Foundation) within the Priority Program SPP 2244 (535377524) and from the European Research Council (ERC) under the European Union’s Horizon 2020 research and innovation programme (grant agreement No. 820254).  A.F, L.K and D.M.K acknowledge funding by the Deutsche Forschungsgemeinschaft (DFG, German Research Foundation) under RTG 1995 and within the Priority Program SPP 2244 ``2DMP''. F.L. acknowledges support by the Austrian Science Fund (FWF), project I-3827-N36 and by the COST association, COST action CA18234. This work was supported by the Max Planck-New York City Center for Nonequilibrium Quantum Phenomena.
Fabrication of the samples was supported by the Helmholtz Nano Facility~\cite{Albrecht2017}. 
K.W. and T.T. acknowledge support from JSPS KAKENHI (Grant Numbers 19H05790, 20H00354 and 21H05233).
\end{acknowledgements}
~\\

R.J.D. and A.R. performed the experiments. A.R. fabricated the samples, L.W. built the laser cutting setup; and K.W. and T.T. grew the hBN crystals. R.J.D. performed the data analysis and the results were interpreted with the input of A.R., A.F., L.K., L.W., D.M.K., F.L., B.B. and C.S. R.J.D wrote the manuscript with the input of A.R., A.F., L.K., L.W., D.M.K., F.L., B.B. and C.S.

\appendix
\setcounter{equation}{0}
\setcounter{figure}{0}
\setcounter{table}{0}
\makeatletter
\renewcommand{\theequation}{A\arabic{equation}}
\renewcommand{\thefigure}{A\arabic{figure}}
\renewcommand{\thesection}{A\arabic{section}}   
\renewcommand{\thetable}{A\arabic{table}}

\section{Fermi surface area extracted from the quantum oscillations} \label{sec:appBfield}
\begin{figure*}
    \centering
    \includegraphics{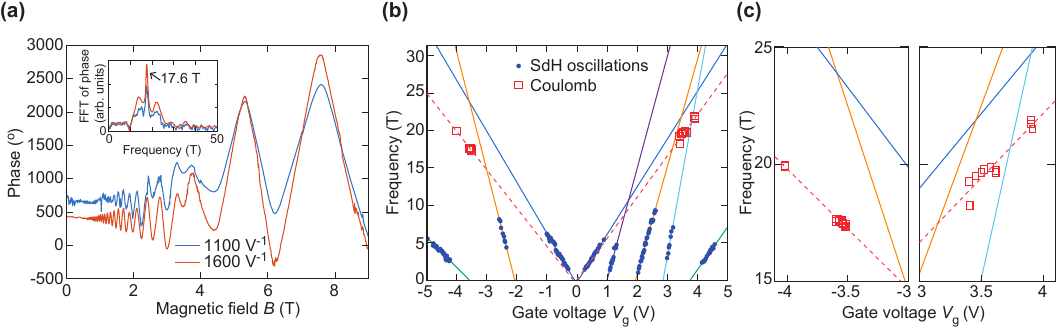}
    \caption{(a) Phase of 2 prominent frequency components extracted from the fast Fourier transform in Fig.~\ref{fig:Bfield}(b). 
    (b) Quantum oscillation frequency of the magnetoresistance oscillations (see Supplemental Material~S1~\cite{Supplemental}) compared to the quantum oscillations in the phase of the Coulomb oscillations, as a function of gate voltage. 
    (c) Zoom-in of the quantum oscillation frequency in the case of the Coulomb oscillation.}
    \label{fig:Bfieldappendix}
\end{figure*}
The electrostatically-induced shifts in the position of the Coulomb resonances provides important clues about the transport characteristics between the metal leads and the confined regions. 
We plot the phase of the Coulomb oscillations (which is proportional to the Coulomb resonance position) obtained from the FFT analysis of the two prominent frequency components [Fig.~\ref{fig:Bfield}(b)] in Fig.~\ref{fig:Bfieldappendix}(a). 
This reveals a quantum oscillation that is periodic in $1/B$, and the quantum oscillation frequency can be extracted from an additional FFT analysis of the phase [inset of Fig.~\ref{fig:Bfieldappendix}(a)]. 
By monitoring the phase of the oscillations at different gate voltages, we can extract the quantum oscillation frequency and plot these as red squares in Figs.~\ref{fig:Bfieldappendix}(b)--\ref{fig:Bfieldappendix}(c). 
Since these quantum oscillations have the same microscopic origin as the Subnikov-de Haas (SdH) oscillations observed in the sample's bulk magnetoresistance (see Supplemental Material~S1~\cite{Supplemental}), we also add the SdH frequency as a comparison [blue dots in Fig.~\ref{fig:Bfieldappendix}(b)].
The quantum oscillation frequencies obtained from the Coulomb oscillations form a straight line that intersects the charge neutrality point at zero magnetic field in Fig.~\ref{fig:Bfieldappendix}(c). 
In contrast, the SdH frequency intersects with fractional superlattice fillings at zero magnetic field due to the Dirac revival effect in tBLG, which reconstructs the Fermi surface area at fractional fillings~\cite{zondiner2020cascade}.
We observe the same disparity in a second device, labeled D4, which is presented in the Supplemental Material~S7~\cite{Supplemental}. 
At higher gate voltages, we note that the lever arm of the gate decreases, which explains why the quantum oscillation frequency from the Coulomb oscillations exhibits a more gradual slope [red dashed lines in Figs.~\ref{fig:Bfieldappendix}(b)--\ref{fig:Bfieldappendix}(c)] compared to the SdH frequency component emerging from the charge neutrality point [blue lines in Fig.~\ref{fig:Bfieldappendix}(b)].

The Onsager relation establishes a direct proportionality between the frequency of quantum oscillations and the extremal Fermi surface area in momentum space~\cite{onsager1952interpretation}.
In proximity to an energy gap, it logically follows that both the Fermi surface area should approach 0~nm$^{-2}$ and the frequency of quantum oscillations approaches $0$~T.
Therefore, the fact that the quantum oscillation frequency forms a linear relationship emanating from the charge neutrality point (CNP) is evidence for the absence of moiré-induced energy gaps for the charge carriers that are involved in the tunneling processes in- and out of the charge islands.

\section{Ground state energy of a Wigner crystal in a moir\'e superlattice}
\label{sec:appbeta}
In this Appendix we show that a periodic moir\'e potential may lead to a lowering of the ground state energy of a Wigner crystal. 
Due to the periodic moir\'e potential, a fraction of charge carriers in the triangular Wigner lattice can obtain an even lower energy within the potential, lowering the ground state energy even further than compared to a uniform background.
If we assume a potential with minima $\Delta E$, we can write this additional interaction energy contribution as:
\begin{equation}
    E_{\rm moir\acute{e}} = - A n \Delta E \mathcal{R},
\end{equation}
where $\mathcal{R}$ represents the effective fraction of charge carriers that are in the optimal position to profit from the moir\'e potential.
For simplicity, we assume that $\mathcal{R} = a_{\rm moir\acute{e}}/a$ (where $a_{\rm moir\acute{e}}$ is the size of the moir\'e supercell) such that $\mathcal{R} = 1$ if $a_{\rm moir\acute{e}} = a$. 
Since $1/a = \sqrt{n}$, the moir\'e energy contribution becomes:
\begin{equation}
    E_{\rm moir\acute{e}}  = - A n^{3/2} \Delta E a_{\rm moir\acute{e}},
\end{equation}
this leads to an additional capacitance contribution:
\begin{equation}\label{eq:moireenergy}
    \frac{1}{C_{\rm moir\acute{e}}} = - \frac{3 \Delta E a_{\rm moir\acute{e}}}{4 e^2 A \sqrt{n}}.
\end{equation}
with the same dependency on area and density as the interaction energy for a uniform background.

To estimate how much the ground state energy is lowered, we compare to a tight-binding model of the moir\'e superstructure~\cite{TB_model_moire2022}. For small twist angles, the local rotation between unit cells in the top and bottom layers can be neglected in favor of only considering a rigid displacement vector $\mathbf{d(\mathbf{r})}$ between the unrotated top and bottom layer. In the small angle approximation, ${\mathbf{d(\mathbf{r})}}$ can then be explicitly written as 
$ \mathbf{d}(\mathbf{r}) \approx-\theta \hat{z} \times \mathbf{r}$. All displacements $\mathbf{d}(\mathbf{r})$ lie within the unit cell of the pristine lattice. Consequently, one can map each local configuration at a point $\mathbf{r}$ of the moir\'e supercell to a rigid displacement ${\mathbf{d(\mathbf{r})}}$ in so-called configuration space mapped on the unit cell of the pristine lattice. Following the model we have outlined in~\cite{TB_model_moire2022}, we map out tight-binding parametrizations in configuration
space using a $10\times 10$ grid in configuration space. We use a continuum elasticity model first suggested by Nam et al.~\cite{Nam2017lattice} to calculate the effects of lattice reconstruction in tBLG. From our parametrization we extract the variations in on-site potential at a twist angle of $\approx 1$ degrees, yielding $\Delta E \approx 14$ meV and $a_{\rm moir\acute{e}} = 14$ nm.
From this analysis, we find that the new correlation strength parameter $S = 0.16$, or an increase of 33\% compared to the Wigner crystal with a uniform charge background.
This result gives an idea of the order of magnitude of the correction, but is likely an underestimation of the correction for two reasons.
First, the interaction with the positive background is not included, and will also result in an additional contribution due to the non-uniform charges of the underlying lattice. 
Second, we do not include effects of elasticity in the Wigner crystal, the Wigner lattice will likely deform to ensure a higher fraction is in the optimal position. 

\section{Suppression of kinetic energy in remote bands}\label{sec:bandstructure}
\begin{figure}
    \centering
    \includegraphics{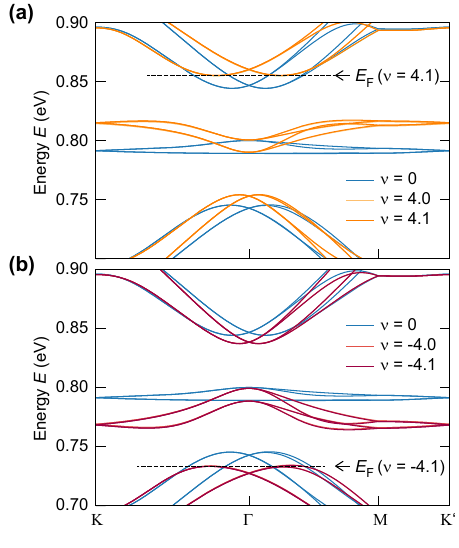}
    \caption{Band structure of 1.018$^{\circ}$ tBLG in the presence of long-ranged electron-electron interactions. Hartree corrections render the bandstructure of tBLG filling-dependent, which leads to a pinning of the van Hove singularitiy to the Fermi energy $E_{\rm F}$ and additional band flattening at the tip of the remote valence and conduction bands.}
    \label{fig:bandflattening}
\end{figure}

The atomic and electronic structure of tBLG is captured within an atomistic modeling approach~\cite{trambly2010localization,koshino2018maximally,fischer2021spin} that  relies on commensurate moiré unit cells with twist angle
\begin{equation}
\cos\theta=\frac{n^{2}+4 n m+m^{2}}{2\left(n^{2}+n m+m^{2}\right)},
\end{equation}
where $m,n \in \mathbb{N}$. 
For the simulations, we use $(n,m)=(32,33)$ corresponding to a twist angle of $\theta = 1.018^{\circ}$ with $N=12'676$ carbon atoms per moiré unit cell. 
The positions of the carbon atoms are relaxed using classical force fields as outlined in Ref.~\cite{klebl2021importance}.
The electronic structure is modeled by a Slater-Koster tight-binding model of the carbon $p_z$-orbitals using the parametrization adopted in Ref.~\cite{koshino2018maximally}.
Near the magic-angle, long-ranged Coulomb interactions were shown to significantly renormalize the single-particle flat bands of tBLG~\cite{Guinea2018,goodwin2020hartree,Rademaker2019,Cea2019} if the system is filled with electrons (holes), which can be captured within a self-consistent Hartree theory. 
Ref.~\cite{goodwin2020hartree} demonstrated that within atomistic modeling approaches, the Hartree potential can effectively be parameterized by an on-site term of the form 
\begin{equation}
V^H(\bvec r) = V_0 \nu \sum_{j} \cos(\bvec G_j \cdot \bvec r), 
\label{eq:atomistic_hartree_potential}
\end{equation}
where $\nu$ denotes the electronic filling $\nu = -4 \dots 4$ of the flat bands with respect to charge neutrality $(\nu=0)$ and $\bvec G_j$ are the three non-equivalent moiré reciprocal lattice vectors that differ by rotations around 120$^{\circ}$. 
The value of the Hartree potential $V_0$ was found to be $V_0 = 5$ meV for the unscreened Coulomb interaction~\cite{goodwin2020hartree, goodwin2020critical}. \\

The (filling-dependent) bandstructure of 1.018$^{\circ}$ tBLG along the high symmetry path $K-\Gamma-M-K'$ is shown in Fig.~\ref{fig:bandflattening}. 
At half-filling, the systems features a set of flat bands that are well separated from the remote valence and conduction bands by a well-defined energy gap. 
Filling the flat bands of tBLG with electrons (upper panel) or holes (lower panel), shifts the energies at the $K,K'$ points to higher (lower) energies due to the Hartree potential. 
Therefore, the flattest sections of the bands follow the Fermi energy $E_F$, which leads to a pinning of the van Hove singularities~\cite{Cea2019}. 
Furthermore, the Hartree potential affects the flatness of the tip of the valence (conduction) band manifold as indicated by the dashed line at filling factor $\nu = \pm 4.1$. 
This effect reduces the kinetic energy of charge carriers at the edge of the remote band, which may account for the prominent observation of a negative compressibility in this regime. 

\section{Methods}\label{sec:methods}
\subsection{Samples}
\textbf{Exfoliation:} The flakes used for fabrication were mechanically exfoliated onto a silicon wafer with 90 nm thick thermally grown silicon dioxide~\cite{Novoselov2005Jul}.
Graphite flakes ("graphenium") where obtained from NGS Naturgraphit GmbH.

\textbf{Stacking:}  Device D1 was fabricated using the stack-and-tear method~\cite{cao2016superlattice}.
For Device D1, we used polyvinyl alcohol (PVA) and polydimethylsiloxane (PDMS), and the stacking of the flakes was performed using the parameters described in Ref.~\cite{schapers2022raman}. 
 For samples D2, D3, D5 and D6 we used a polybisphenol A carbonate (PC) stamp on top of a PDMS stamp~\cite{wang2013one}. In addition, the single-layer graphene flakes of these devices were pre-cut using a laser. 
Device D4 was produced using a poly bisphenol a carbonate (PC) stamp on PDMS~\cite{wang2013one}.

For each sample, the thicknesses of the flakes are measured in tapping mode atomic force microscopy and the results are shown in Table~\ref{tab:thickness}.
\begin{table}[h!]
\caption{Thicknesses of the flakes used in each sample. Each thickness was determined by measuring the step-height at the edge of the flake after fabrication in an atomic force microscope. The uncertainty on each thickness is 1~nm.} \label{tab:thickness}
\begin{ruledtabular}
\begin{tabular}{cccc}
Sample & Bottom hBN (nm) & Top hBN (nm)   & Graphite (nm)  \\ \hline
D1     & 29  & 32  & 8   \\
D2      & 32   & 24  & 5   \\
D3      & 32  & 26  & 4  \\
D4      & 35  & 37  & 10 \\
D5      & 31 & 22 & 2 \\
D6 & 22 & 22 & 3
\end{tabular}
\end{ruledtabular}
\end{table}

\textbf{Laser cutting:} Laser cutting was performed using a focused supercontinuum laser, coupled into an optical microscope setup. 
The output of the laser was spectrally filtered to contain wavelengths of 400-550~nm, which maximises the ratio of absorption in graphene compared to the absorption in Si for the used oxide thickness of 90~nm. 
The pulse duration on the sample is estimated to be few 10s of picoseconds. 
The laser has a maximum repetition rate of 20~kHz, but is typically operated at 4~kHz. 
The power can be varied using absorption filters, with a higher power increasing the width of the cut. 
Typical pulse energies used for laser cutting are 4~nJ, focused down to a submicrometer spotsize, equivalent to a maximum intensity of approximately $2\times 10^{10}$ W/cm$^2$.

\textbf{Fabrication:}
For fabrication, we used a 50K/950K polymethyl methacrylate (PMMA) double-layer as our resist system (Allresist 631.09 and 679.04 both spin-coated at 4000 rpm and baked at 150$^{\circ}$ for 3 minutes per layer, with a mixture of 3 parts isopropylalcohol and 1 part water as a developer) and e-beam lithography (Vistec EBPG5200+, 100 keV, clearance dose of 500 $\mu$C/cm$^2$). 
The Hall-bar device was fabricated by first patterning holes into the top hBN layer.
These holes were etched in a reactive ion etcher by first using a short oxygen plasma step (Oxford PL 100 / ICP at 20 W RF power, 40 sccm, as low pressure as possible $\sim$ 8 $\mu$bar for $\sim$5 s) to remove contamination, followed by a CF$_4$ plasma etch (10 W, 40 sccm, low pressure), which significantly slows down at the graphene layer~\cite{Uwanno2018Aug}. 
The duration of the CF$_4$ plasma step was adjusted to the top hBN thickness to prevent over-etching into the bottom hBN layer.
This was followed by another brief oxygen plasma step to remove the graphene, after which chrome and gold were deposited using e-beam evaparation, followed by a lift-off in warm acetone (without sonication).
This results in a clean one-dimensional edge contact to the tBLG~\cite{wang2013one}.
Subsequently, the Cr/Au metal contacts and bond pads were patterned and evaporated using the same lift-off process.

For device D4, the fabrication was performed using a two-step process, where first the device geometry was structured using the CF$_4$ plasma, and the electrodes were deposited in the second step. 
The contact resistances in this process are significantly higher than in the other devices, and this fabrication approach was was not pursued further. 

\subsection{Twist angle determination}
\textbf{Device D1} To determine the twist angle in the 750~nm constriction (C1), we extract the superlattice filling $n_s = \num{2.43e12} \pm \num{0.04e12}$~cm$^{-2}$ from the Landau fan (Supplemental Material~S1~\cite{Supplemental}), and use the equation $n_s = {8\theta^2}/{\sqrt{3} a_{l}^2}$~\cite{cao2018correlated} (where $a_l = 0.246$ nm is the lattice constant of graphene) to find the twist angle in $\theta = 1.022^{\circ}$.
For the remaining constrictions, we used the position of the band insulating features to determine the twist angle and find twist angles of 1.07$^{\circ}$, 0.97$^{\circ}$ and 0.91$^{\circ}$ for constrictions C2, C3 and C4, respectively. 

\textbf{Device D2} consists of a 1~$\mu$m-wide Hall bar geometry with a twist angle estimated from the Landau fan to be 0.97$^{\circ}$. 
Additionally, we found an additional alignment between one of the graphene layers and the hBN, with a twist angle of $\sim$ 0.7$^{\circ}$. 
We found no effects from this additional alignment on the charging spectra studied in this work. 
An optical image of this device, the conductance traces and power spectra of this device are presented in the Supplemental Material~S4~\cite{Supplemental}.

\textbf{Device D3} is a similar 1~$\mu$m-wide Hall bar geometry without constrictions. The device broke down during the Landau fan measurement, therefore we could only extract the lever arm of the back gate, but not the position of the additional Landau fan. 
From the position of the insulating states, we are nevertheless able to estimate the twist angle to be 1.14$^{\circ}$. 
An optical image of this device, the conductance traces and power spectra of this device are also presented in the Supplemental Material~S4~\cite{Supplemental}.

\textbf{Device D4} due to high contact resistances, it was not possible to extract the lever arm or the twist angle from magnetotransport experiments. 
Therefore, the twist angle was estimated using a parallel-plate capacitor model and the position of the insulating states. 
We estimate the twist angle to vary between 1.29$^{\circ}$ and 1.45$^{\circ}$ on this sample. 
Sample imaging and measurement results are presented in Supplemental Material~S7~\cite{Supplemental}.

\textbf{Device D5} on this device, we did not obtain a clear additional Landau fan, but the lever arm could be estimated from the charge neutrality point. 
The twist angle was estimated from the position of high resistance features found in transport measurements, which occur at $\nu = \pm 8$ in the low-twist-angle regime of this device~\cite{cao2018correlated}. 
We find a twist angle of 0.65$^{\circ}$, and the results on this device are presented in Supplemental Material~S5~\cite{Supplemental}.

\textbf{Device D6} This device incorporated an additional WSe$_2$ (HQ graphene) into the stack which was picked up after picking up the tBLG.
The tBLG graphene was relaxed back to near-Bernal stacking, since no evidence of moir\'e induced satellite peaks was found. 
This device serves as a control sample, which is presented in the Supplemental Material~S6~\cite{Supplemental}.

 \subsection{Experimental setup}
Electrical characterization of all samples was performed in a $^3$He/$^4$He wet dilution  refrigerator (Oxford KelvinoxMX400) with a base temperature of 32.5 mK.
All electrical wiring is directly connected to the sample with only a 1 kOhm pre-resistor in the BNC connector box used to connect to the instruments.
The home-build amplifiers and IV converters are each placed in a shielded box outside the refrigerator.
The total resistance between the BNC connector box to the sample holder is 1.24 k$\Omega$ for each line at room temperature. 
An out-of-plane magnetic field is applied with a superconducting magnet mounted in the liquid helium bath. 
We locate the sample inside the coil of the magnet while avoiding magnetic materials in the insert to ensure a uniform magnetic field. 

\subsection{Data aquisition and analysis}
\textbf{Measurement of temperature-dependent resistance:} The temperature-dependent resistance in Figs.~\ref{fig:D1tempdep}(d)--\ref{fig:D1tempdep}(e) was measured using a homebuild IV-converter with a gain of $\num{10e6}$ connected to two side contacts. 
This IV converter also applied a symmetric AC bias voltage to these contacts with an amplitude of 100~$\mu$V (RMS) through a 1/10,000 voltage divider.
The excitation signal was supplied by a Stanford SR830 lockin amplifier operating at 69.6~Hz and this apparatus also detected the signal from the IV converter.
On the opposite side-contact pair, a differential amplifier with a gain of 1000 is connected to measure the voltage drop and this signal is measured with a second lockin amplifier. 
A voltage source (Yokogawa 7651) was connected to the gate through a 1~M$\Omega$ resistor. 
The measurement was performed during the condensing of the mixture, circulating of the mixture and cooldown to base temperature of the dilution refrigerator. 
During this procedure the gate voltage was constantly swept and at each data point the temperature on the mixing chamber plate is recorded.
During the cooldown procedure the mixing chamber rapidly cools once the condensed mixture enters, therefore reliable measurements of the temperature between $\sim 2.5$ K and $\sim 4.5$ K were not possible.
The resulting data is plotted on a meshed grid with interpolation using the Matlab 'pcolor' function.

\textbf{2-point conductance measurements:} The conductance traces used to construct the power spectra are obtained by a 2-point measurement of the conductance using the homebuild IV converter and an AC bias of 100 $\mu$V RMS at a frequency of 69.6 Hz.
The voltage source for the gate (Yokogawa 7651) was set to a range of 10~V, giving a resolution of 100~$\mu$V, and this is supplied to the system through a twisted-shielded cable with a low-pass filter with a cut-off frequency of 1.6 kHz to reduce output noise.
The resolution of 100 $\mu$V is also the step size at which the gate voltage is swept, resulting in a Nyquist frequency of 5000 V$^{-1}$.
For the bias spectroscopy data in Fig.~\ref{fig:biasdependence}, an additional voltage source was used to apply a symmetric DC bias.

\textbf{Determination of the power spectra:}
To calculate the power spectra $P_{\omega}$, we first perform a windowed autocorrelation on the gradient of the conductance with the Matlab function {corrgram}~\cite{corrgram}.
We use a window of 200 samples, with an maximum lag of 200 points and overlap between the windows of 90\%.
Note, that with this window we can only resolve frequency components $f_g > 50$ V$^{-1}$.
We then calculate the Fourier transform of the autocorrelation function using the Matlab implementation 'ezfft'~\cite{ezfft}.
By the convolution theorem this approach is equivalent to calculating $|\mathcal{F}\{\mathrm{d} G/ \mathrm{d} V_g\}|^2$.

\textbf{Fitting to the power spectra:}
To fit Eq.~\eqref{eq:slopeequation} to our data, we find the maxima in the power spectra at each gate voltage.
Due to the limited frequency resolution, this can lead to multiple gate voltage values having the same capacitance in Fig.~\ref{fig:fits}, but this is not a problem for the least squares fitting procedure.
We then load this into the curve fitting tool 'cftool' in Matlab.
We then manually select the correct maxima that belong to a single frequency component, and fit Eq.~\eqref{eq:slopeequation}.
cftool automatically calculates the 95\% confidence interval, which defines the error in the fitting results presented in this work. 
We find the hBN thickness $d$ from AFM scanning the stack before fabrication, their values are shown in Table~\ref{tab:thickness} under the column: "bottom hBN". 
The lever arm $\alpha$ is extracted during the extraction of the twist angle, as described above. 

\textbf{Magnetic field dependence:}
For Figs.~\ref{fig:Bfield}(b) the fast Fourier transform was calculated at each magnetic field using the standard FFT implementation in Matlab.
This fast Fourier transform gives a complex number, of which the angle can be calculated to obtain the phase.
The resulting phase was unwrapped to obtain the continuous signal shown in Fig.~\ref{fig:Bfieldappendix}(a).
To find the frequency of the phase in $1/B$ as shown in the insert of Fig.~\ref{fig:Bfieldappendix}(a), the phase signal was interpolated using a spline interpolation on a grid that is equally spaced in $1/B$. 
After this, the FFT can be calculated to find the quantum oscillation frequency.
The resulting frequencies are shown as red squares in Figs.~\ref{fig:Bfieldappendix}(b)--\ref{fig:Bfieldappendix}(c).
The same interpolation approach is taken on the magnetoresistance data (presented in Supplemental Material~S1~\cite{Supplemental}).
After this, we find the maxima at each B-field and manually select the correct frequency components, which are shown as blue dots in Fig.~\ref{fig:Bfieldappendix}(b).

\clearpage

\onecolumngrid

\renewcommand{\theequation}{S\arabic{equation}}
\renewcommand{\thefigure}{S\arabic{figure}}
\renewcommand{\thesection}{S\arabic{section}}   
\renewcommand{\thetable}{S\arabic{table}}
\setcounter{equation}{0}
\setcounter{figure}{0}
\setcounter{table}{0}

\section*{Supplemental Material}

\section*{S1: Additional bulk transport data on the 750 nm constriction}
\subsection*{S1.1: Fragile superconductivity in the 750 nm constriction}\label{sec:supercond}
\begin{figure*}[h!]
    \centering
    \includegraphics{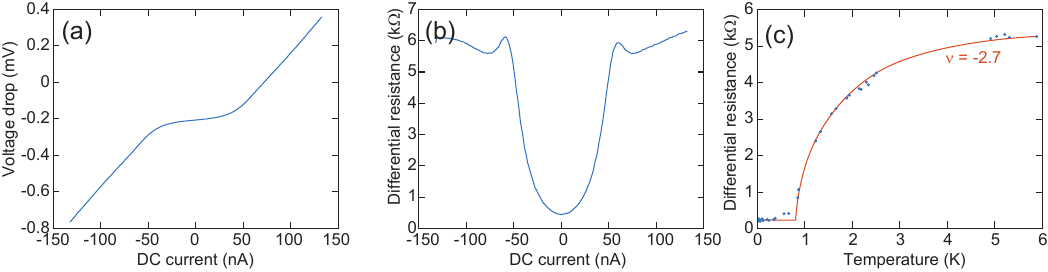}
    \caption{Observation of fragile superconductivity in constriction C1. (a) Temperature dependence of the 4-point resistance, showing correlated insulators (CI), band insulators (BI) and the fragile superconducting dome (SC). (b) DC voltage drop and (c) differential resistance measured over the 750 nm constriction as a function of DC current. (d) Temperature dependence of the resistance at a filling factor $\nu = -2.7$. Red solid line is a guide to the eye.  }
    \label{fig:differential_conductance}
\end{figure*}
In Figs.~1(d)-1(e) of the main text we show the 4-point resistance measured in constriction C1 as a function of temperature. 
Besides band insulators (BI) and correlated insulators (CI), we find a region with low resistance near $\nu = -2.7$ at low temperatures, resembling a superconducting dome that is often observed in magic angle tBLG. 
To investigate this further, we measured the voltage-current characteristic at a backgate voltage of -2.43~V and base temperature.
Figure~\ref{fig:differential_conductance} shows the DC voltage drop [Fig.~\ref{fig:differential_conductance}(a)] and differential resistance [Fig.~\ref{fig:differential_conductance}(b)] as a function of DC current. 
DC signals were simultaneously obtained with AC signals (to obtain the differential resistance) using standard lockin techniques at a frequency of 19.11~Hz.
We observe a plateau in the voltage drop around zero DC current and two maxima in the differential resistance. 
Both are characteristic features of superconductivity, however, a 400~$\Omega$ resistance remains.
The temperature-dependence of the differential resistance shows a characteristic drop as the temperature decreases, below which it shows a plateau at a finite resistance [Fig.~\ref{fig:differential_conductance}(c)]. 
Therefore, we conclude that we do not observe robust superconductivity as often observed in tBLG  \cite{park2022robust}, but rather a fragile superconducting dome. 
This may be attributed to a resistance in series to a superconducting region in the 750 nm constriction.
Alternatively, the rather large voltage offset of the differential amplifier of 0.2 mV [Fig.~\ref{fig:differential_conductance}(a)] might also be responsible for breaking the superconducting state, if this is indeed an offset on the input of the amplifier. 

\subsection*{S1.2: Chern insulators in the 750 nm constriction}\label{sec:chern}
\begin{figure*}[h!]
    \centering
    \includegraphics[]{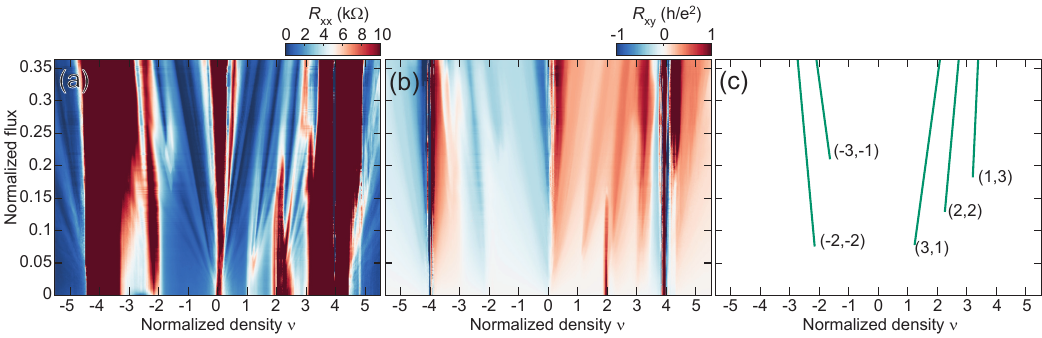}
    \caption{Observation of Chern insulators in constriction C1. (a) Longitudinal resistance as a function of normalized density $\nu = n A_{\rm uc}$ and normalized flux $\Phi/\Phi_0$. (b) Transverse resistance in units of $h/e^2$. (c) Wannier diagram of observed Chern insulators}
    \label{fig:chern}
\end{figure*}

Figure~\ref{fig:chern} shows the longitudinal ($R_{\rm xx}$) and transverse ($R_{\rm xy}$) resistance as a function of normalized density and flux. 
These normalized quantities were found by calculating the supercell area from the superlattice density. 
This directly gives the normalized density, which is equivalent to the number of charge carriers per supercell. 
The flux through one supercell is normalized with respect to the flux quantum. 

As shown in Fig.~\ref{fig:chern}(c) we observe a series of Chern insulators, whose topological invariants are $(s,t)$, where $s$ is the slope and $t$ is the offset. These states are only included in Fig.~\ref{fig:chern}(c) if $R_{xy} = 1/s ~ [h/e^2]$. They have been observed with the same topological invariants in other works in tBLG near the magic angle \cite{nuckolls2020strongly,tomarken2019electronic,stepanov2021competing,saito2021hofstadter,wu2021chern,park2021flavour} and arise from the energy spectrum of the circular orbits of the charge carriers in the periodic potential of the moiré superlattice. 

\clearpage

\section*{S2: Dependence of the Coulomb oscillations on the sample width}
\begin{figure*}[h!]
    \centering
    \includegraphics{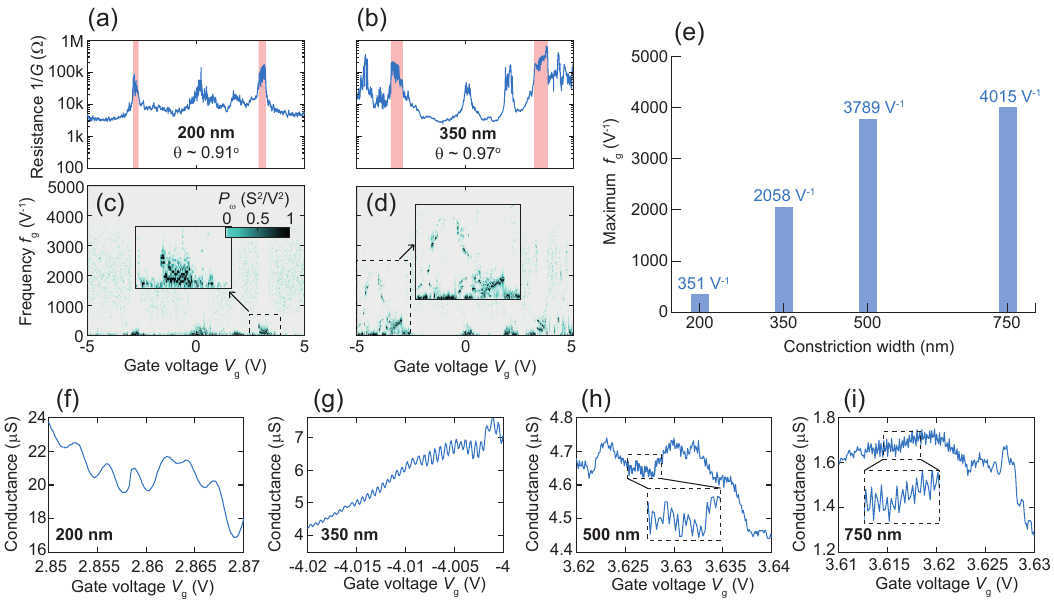}
    \caption{(a),(b) 2-point conductance traces over the full gate voltage range for the 350 and 200 nm wide constriction, respectively. The band insulators (BI) are highlighted by red shaded regions. 
    (c)--(d) Power spectra $P_{\omega}$ as a function of gate voltage for the 350 and 200 nm wide constriction, respectively.
    (e) The highest discernible periodic oscillation frequency as a function of constriction width.
    (f-i) Examplary conductance traces in a 20~mV window for the 200, 350, 500 and 750 nm wide constriction, respectively. }
    \label{fig:sizedep}
\end{figure*}
To elucidate the location on the sample where these confinements are formed, we analyze conductance traces taken over each constriction in device D1. 
Two example traces on the 200~nm and 350~nm wide constriction are shown in Figs.~\ref{fig:sizedep}(a),~\ref{fig:sizedep}(b), respectively. 
To qualitatively evaluate the frequency content within each constriction, we examine the power spectra of each in Figs.~\ref{fig:observe_oscillations}(b),~\ref{fig:sizedep}(c),~\ref{fig:sizedep}(d) (for the 500 nm wide constriction, see section S4), identifying the highest confidently discernible periodic oscillation frequency in each case.
This analysis reveals a diminishing trend of high-frequency contributions with decreasing sample width, as visually depicted in Figs.~\ref{fig:sizedep}(e).
Specifically, at a sample width of 200 nm, no high-frequency Coulomb oscillations are evident [Fig.~\ref{fig:sizedep}(c)], whereas at a width of 350~nm, distinct periodic components are still observable up to 2058~V$^{-1}$ [Fig.~\ref{fig:sizedep}(d)].
To further illustrate the frequency content within each constriction, we present exemplary traces of the conductance within a 20~mV gate voltage span (close to the maximum in frequency), showcased in Figs.~\ref{fig:sizedep}(f)--(i).  

\clearpage
\section*{S3: Fitting results to determine S}
\begin{table*}[h!]
\centering 
\caption{Fitting results on the frequency components with $S$ as a fitting parameter. }
\begin{ruledtabular}
\begin{tabular}{ccccccc}
Component & Near filling & $\sqrt{A}$       & $V_0$               &$S$ &  $R^2$ \\ \hline
C1.1        & -4           & 307.4  $\pm$ 12.0 & -3.909 $\pm$ 0.047  & $\num{1.09} \pm \num{0.13}$  & 0.986 \\
C1.2        & -4           & 126.2 $\pm$ 10.7 & - 3.559 $\pm$ 0.118 &  $\num{0.90} \pm \num{0.30}$  &  0.911 \\
C1.3        & +4           & 341.8 $\pm$ 24.3 & 3.600 $\pm$ 0.040   &  $\num{0.57} \pm \num{0.17}$ &  0.987 \\
C1.4        & +4           & 162.7 $\pm$ 12.5 & 3.834 $\pm$ 0.016   &  $\num{0.48} \pm \num{0.12}$ &  0.950 \\
C2.1        & -4           & 259.6 $\pm$ 31.0 & -3.605 $\pm$ 0.378  &  $\num{2.09} \pm \num{0.60}$ &0.974 \\
C2.5        & +4           & 153.9 $\pm$ 13.0 & 3.797 $\pm$ 0.021   & $\num{0.62} \pm \num{0.13}$ &0.946 \\
C3.2       & -4           & 342.1 $\pm$ 44.5 & -4.166 $\pm$ 0.102  &  $\num{0.69} \pm \num{0.35} $ & 0.985 \\
D2.1        & -4           & 119.0 $\pm$ 12.7 & -4.653 $\pm$ 0.052  &  $\num{1.21} \pm \num{0.23}$ &  0.977 \\
D2.2        & -4           & 201.0 $\pm$ 22.4 & -3.968 $\pm$ 0.075  &  $\num{0.92} \pm \num{0.30}$  & 0.943 \\
D2.3        & -4           & 231.5 $\pm$ 70.5 & -5.431 $\pm$ 0.861  &  $\num{2.70} \pm \num{1.49}$ &  0.963 \\
D2.4        & +2           & 301.9 $\pm$ 7.9 & 2.758 $\pm$ 0.070   &   $\num{1.05} \pm \num{0.16}$ & 0.978 \\
D3.3        & +3           & 216.7 $\pm$ 34.0 & 4.276 $\pm$ 0.244   &  $\num{2.50} \pm \num{0.36}$ & 0.974
\end{tabular}
\end{ruledtabular}
\end{table*}

\clearpage

\section*{S4: Remaining plots of the observed frequency components and fits with $S=0.84$}\label{sec:remaining}
\begin{figure*}[h!]
    \centering
    \includegraphics{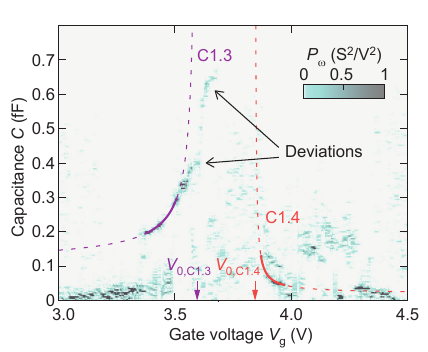}
    \caption{Frequency components observed on the electron-side in the 750-nm-wide constriction (C1) on sample D1 and fits using Eq.~\eqref{eq:slopeequation}.}
    \label{fig:osc750nmremaining}
\end{figure*}

Figure~\ref{fig:osc750nmremaining} shows the frequency components found near the band insulators on the electron side in sample C1. Component C1.3 is interesting, because the capacitance is lower than expected from our model at low densities. While the cause is unknown, it may be related to the large average distance $a$ exceeding the thickness of the hBN $d$, causing a significant change in the dielectric screening. It could also be the combined effect of two distinct charge islands with similar sizes, but different $V_0$. These deviations are not observed in any other frequency component. 

\begin{figure*}[h!]
    \centering
    \includegraphics{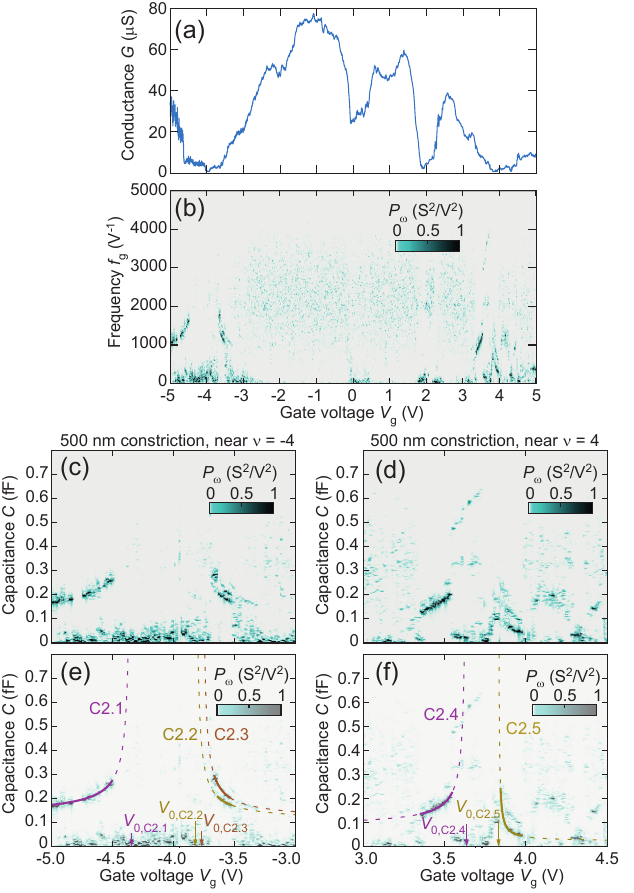}
    \caption{Frequency components observed in the 500 nm constriction (C2). (a) 2-point conductance as a function of gate voltage. (b) Power spectrum as a function of gate voltage and gate-voltage-frequency over the whole gate voltage range. (c) Power spectrum close to the band insulator on the hole-side and (e) fit to 3 frequency components. (d) Power spectrum close to the band insulator on the electron side and (f) fit to two frequency components. }
    \label{fig:osc500nm}
\end{figure*}

Figure~\ref{fig:osc500nm}(a) shows the 2 point conductance $G$ as a function of gate voltage over the 500 nm constriction (C2) in device 1. The band insulators are found at different gate voltages than for the 750 nm constriction. This may be attributed to a slight change in twist angle. From the shift in the position of the band insulator we estimate the twist angle to be $\sim 1.07^{\circ}$. We observe several frequency components near the band insulating features in Fig.~\ref{fig:osc500nm}(b). In Figs.~\ref{fig:osc500nm}(c)--\ref{fig:osc500nm}(f) we identify 5 components that fit well to Eq.~\eqref{eq:slopeequation}. 

\begin{figure*}[h!]
    \centering
    \includegraphics{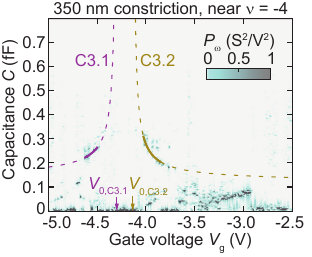}
    \caption{Power spectrum and fit to two frequency components observed in the 350-nm-wide constriction (C3). }
    \label{fig:osc350nm}
\end{figure*}

Figure~\ref{fig:osc350nm} shows the fits for the 350 nm wide constriction (C3) on device 1. In this region, the twist angle is found to be lower than for the 750 nm constriction. Based on the position of the band insulators, we estimate the twist angle in this region to be $\sim 0.97^{\circ}$.

\begin{figure*}[ht!]
    \centering
    \includegraphics{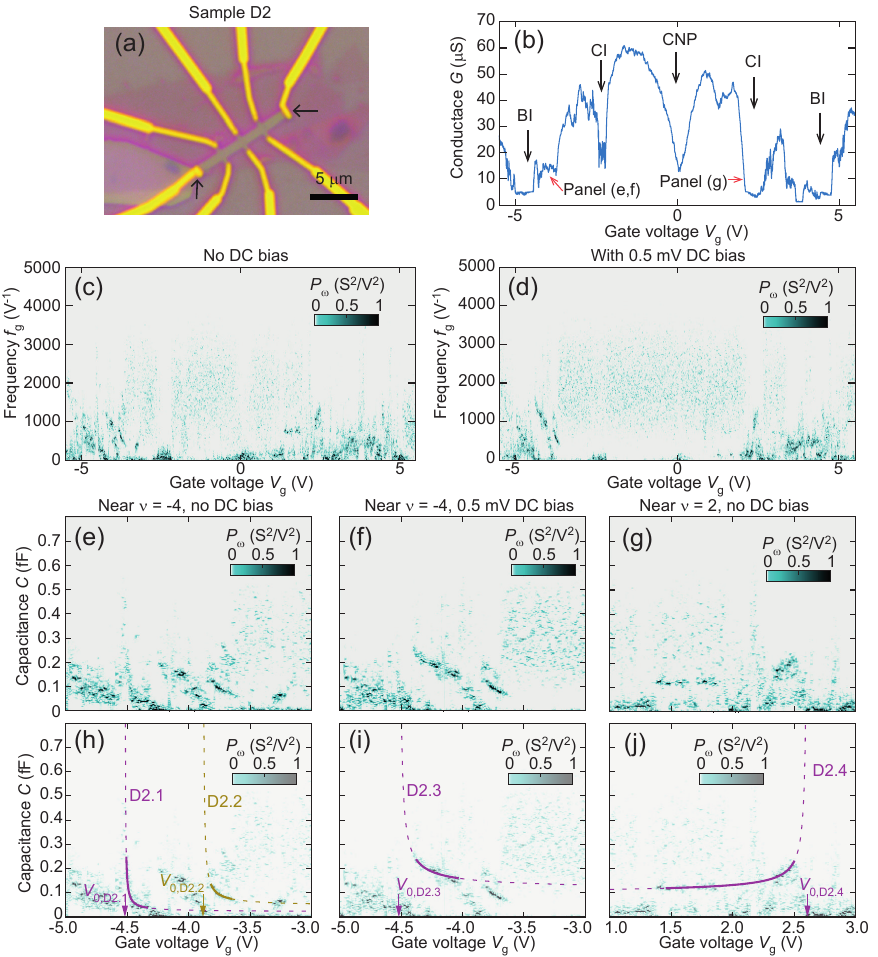}
    \caption{Conductance oscillations observed in device D2. This sample consists of a simple Hall-bar structure (no constrictions) and tBLG with an estimated twist angle of 0.97$^{\circ}$. (a) Optical microscopy image of the device, the two contacts highlighted by black arrows are used for the conductance measurement. (b) Two-point conductance as a function of gate voltage, highlighting the band insulators (BI) and correlated insulators (CI). (c) Power spectrum as a function of gate voltage and frequency without a DC bias applied and (d) with a 0.5 mV DC bias applied. (e) Zoom-in of the Power spectrum as a function of gate voltage and capacitance near the band insulator on the hole-side and (h) including fits to two frequency components. (f) Power spectrum near the band insulator on the hole side with a 0.5 mV bias and (i) fit to an additional frequency component. (g) Power spectrum near half-filling on the hole side and (j) fit to the frequency component.}
    \label{fig:sample2}
\end{figure*}
Figure~\ref{fig:sample2}(a) shows an optical image of sample D2, which consists of a 1~$\mu$m-wide Hall bar with an estimated twist angle of 0.97$^{\circ}$. In the experiment, we measured the conductance along the total length of the device, using the two contacts indicated by black arrows in Fig.~\ref{fig:sample2}(a). We find the frequency components near $\nu = -4$ [Figs.~\ref{fig:sample2}(e),~\ref{fig:sample2}(f),~\ref{fig:sample2}(h),~\ref{fig:sample2}(i)] and near $\nu = 2$ [Figs.~\ref{fig:sample2}(g),~\ref{fig:sample2}(j)] on this sample. Near the band insulator at $\nu = -4$, initially we find two components denoted D2.1 and D2.2 as shown in Figs.~\ref{fig:sample2}(c),~\ref{fig:sample2}(e),~\ref{fig:sample2}(h). To break the insulating state somewhat, we repeated the measurements with an additional 0.5~mV DC bias applied to the sample [Figs.~\ref{fig:sample2}(d),~\ref{fig:sample2}(f),~\ref{fig:sample2}(i)]. This reveals an additional frequency component D2.3 in a region that was previously masked by the insulating state.

\begin{figure*}[ht!]
    \centering
    \includegraphics{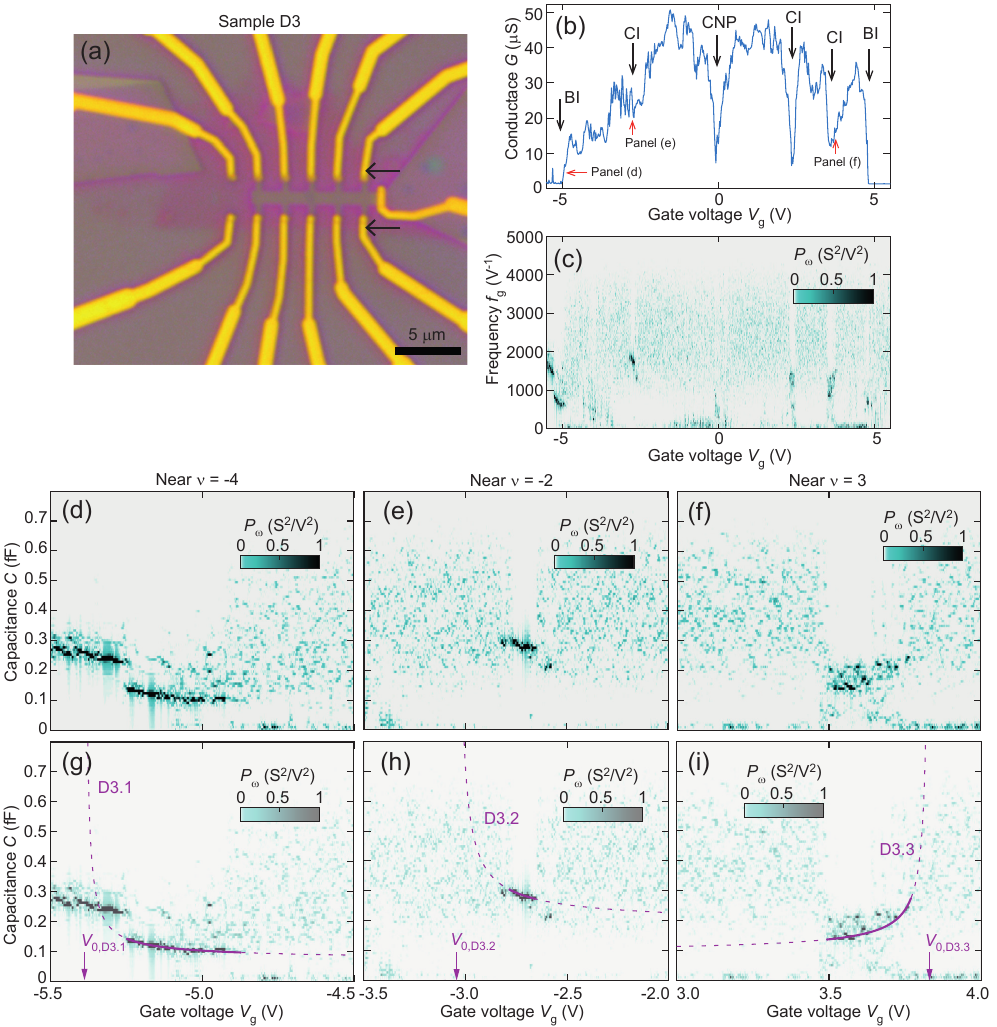}
    \caption{Conductance oscillations observed in sample D3 with an estimated twist angle of 1.14$^{\circ}$. (a) Optical image of the device, showing the two contacts used for 2-point characterization. (b) 2-point conductance as a function of gate voltage. The band insulator (BI), correlated insulators (CI) and charge neutrality point (CNP) are highlighted. (c) Power spectrum obtained from a 4-point measurement (the oscillations are more prominently visible due to oscillations of the contact resistance). (b) Zoom-in near the band insulator on the hole side and (g) fit to the frequency component. (e) Zoom-in near half-filling on the hole side and (h) fit to the component. (f) Zoom-in near 3/4 filling on the electron side and (i) fit to the frequency component. }
    \label{fig:sample3}
\end{figure*}
Sample D3 consists of a Hall bar structure, however only in one of the contacts the graphene was found to be twisted [bottom contact indicated by a black arrow in Fig.~\ref{fig:sample3}(a)], while in the remainder of the sample no moiré-induced satellites are found. When measuring the 2-point conductance [Fig.~\ref{fig:sample3}(b)], we find the band insulators and correlated insulators at fractional fillings. Near the band insulators, we find two distinct frequency components near $\nu = -4$ [Figs.~\ref{fig:sample3}(d),~\ref{fig:sample3}(g)]. We fit only to one component D3.1 [Fig.~\ref{fig:sample3}(c)], because the other component is not fully captured in the gate voltage trace which makes it difficult to find $V_0$ from the fit. During the Landau fan measurement to extract the density, the sample broke, meaning that only data between 7 and 9 T could be used to estimate the lever arm. 

In sample D3, the conductance oscillations are prominently visible in a 4-point measurement if this contact is used as a voltage probe due to the oscillating contact resistance. Due to the larger amplitude, this 4-point measurement is used to detect the oscillations in Fig.~\ref{fig:sample3}. 

In Table~\ref{tab:remain} we summarize the fitting results of all the frequency components found near the band insulators.

\begin{table*}[ht!]
\caption{Summary of the fitting parameters for all frequency components with a fixed value of $S = 0.84$. \label{tab:remain}}
\centering
\begin{ruledtabular}
\begin{tabular}{cccccccc}
Component & Twist angle $\theta$                                 & Lever arm $\alpha$ (m$^{-2}$ V$^{-1}$)  &  Filling factor & Band        & $V_0$ (V)          & $\sqrt{A}$ (nm)  & $R^2$ \\ \hline
C1.1       & \multicolumn{1}{c}{\multirow{4}{*}{1.02$^{\circ}$}} & \multirow{2}{*}{$\num{4.96E+15}$} & -4 & Flat & -3.819 $\pm$ 0.003 & 333.3 $\pm$ 1.1  & 0.983    \\ \cline{1-1} \cline{4-8} 
C1.2        & \multicolumn{1}{c}{}                                   &                                   & -4 & Flat &  -3.536 $\pm$ 0.009 & 128.3 $\pm$ 0.9   & 0.911          \\ \cline{1-1} \cline{3-8} 
C1.3       & \multicolumn{1}{c}{}                                   & \multirow{2}{*}{$\num{5.33e15}$} & 4 & Flat & 3.668 $\pm$ 0.004  & 308.3 $\pm$ 2.1  & 0.983       \\ \cline{1-1} \cline{4-8} 
C1.4        & \multicolumn{1}{c}{}   &                                                                & 4 & Remote & 3.771 $\pm$ 0.005  & 132.3 $\pm$ 2.4  & 0.929 \\ \cline{1-8} 
C2.1        & \multirow{5}{*}{1.07$^{\circ}$}                      & \multirow{3}{*}{$\num{4.96E+15}$} & -4 & Remote & -4.265 $\pm$ 0.008 & 336.9 $\pm$ 1.3  & 0.958          \\ \cline{1-1} \cline{4-8} 
C2.2       &                                                         &                                   & -4 & Flat & -3.897 $\pm$ 0.012 & 301.7 $\pm$ 2.8  & 0.939        \\ \cline{1-1} \cline{4-8} 
C2.3        &                                                         &                                   & -4 & Flat &  -3.845 $\pm$ 0.007 & 313.4 $\pm$ 3.7  & 0.980            \\ \cline{1-1} \cline{3-8} 
C2.4       &                                                         & \multirow{2}{*}{$\num{5.33e15}$} & 4 & Flat & 3.709 $\pm$ 0.006  & 269.3 $\pm$ 2.1  & 0.936          \\ \cline{1-1} \cline{4-8} 
C2.5        &                                                         &                                   & 4 & Remote &  3.759 $\pm$ 0.004  & 136.2 $\pm$ 2.1  & 0.940         \\ \hline
C3.1      & \multirow{2}{*}{0.97$^{\circ}$}                      & \multirow{2}{*}{$\num{4.96E+15}$} & -4 & Remote & -4.219 $\pm$ 0.012 & 347.1 $\pm$ 2.7  & 0.947         \\ \cline{1-1} \cline{4-8} 
C3.2      &                                                         &                                   & -4 & Flat & -4.211 $\pm$ 0.005 & 325.3 $\pm$ 2.0  & 0.985           \\ \hline
D2.1        & \multirow{4}{*}{0.97$^{\circ}$}                      & \multirow{4}{*}{$\num{5.148e15}$} & -4 & Flat & -4.580 $\pm$ 0.002 & 142.6 $\pm$ 1.7   & 0.967           \\ \cline{1-1} \cline{4-8} 
D2.2       &                                                         &                                   & -4 & Flat & -3.945  $\pm$ 0.005 & 207.7 $\pm$ 1.8  & 0.943            \\ \cline{1-1} \cline{4-8} 
D2.3       &                                                         &                                   & -4 & Flat & -4.578 $\pm$ 0.009 & 336.5 $\pm$ 1.9 & 0.940             \\ \cline{1-1} \cline{4-8} 
D2.4       &                                                         &                                   & 2 & Flat & 2.671 $\pm$ 0.006 & 312.6 $\pm$ 0.8 & 0.976             \\ \hline
D3.1       &  \multirow{3}{*}{1.14$^{\circ}$}                                 &\multirow{3}{*}{$\num{5.509e15}$}      & -4 & Flat & -5.444 $\pm$ 0.006 & 263.2 $\pm$ 1.1  & 0.945         \\  \cline{1-1} \cline{4-8} 
D3.2       &                                     &                 & -2 & Flat & -3.101 $\pm$ 0.015 & 430.6 $\pm$ 2.8  & 0.900   \\  \cline{1-1} \cline{4-8} 
D3.3      &                                     &                 & 3 & Flat & 3.889 $\pm$ 0.004 & 299.4 $\pm$ 1.8  & 0.952  
\end{tabular}
\end{ruledtabular}
\end{table*}

\clearpage
\section*{S5: Observation of conductance oscillations near $\nu = \pm 8$ in a 0.65$^{\circ}$ twisted sample}\label{sec:nu8}
\begin{figure*}[ht!]
    \centering
    \includegraphics{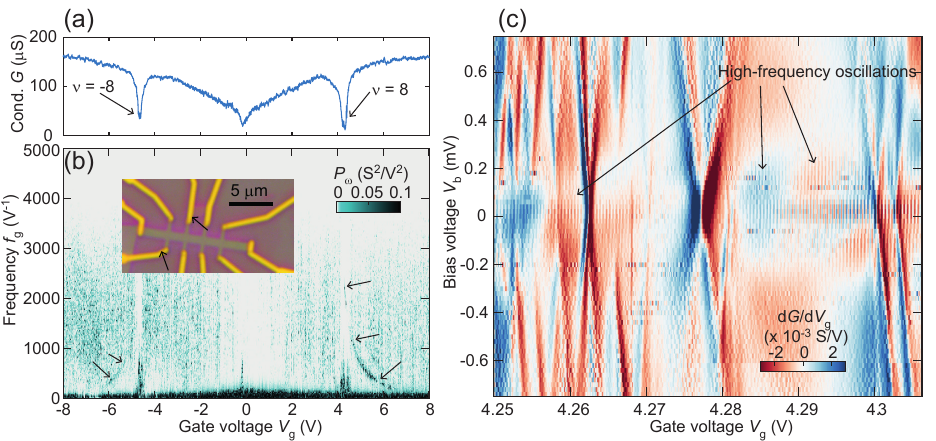}
    \caption{Conductance oscillations observed in sample D5 with an estimated twist angle of 0.65$^{\circ}$. (a) Two-point conductance as a function of gate voltage. At this twist angle, the insulating states are found near $\nu = 8$ \cite{cao2018correlated}, which we identify near $V_g = \pm 4.5$ V. (b) Power spectrum as a function of frequency and gate voltage: Note that the colour scale is a factor 10 lower than in all the other power spectra presented in this work. The inset shows an optical image of the device and the contacts used for 2-point characterization. (c) Bias-dependence of the gradient of the conductance. The typical Coulomb diamond structure is more difficult to discover due to the very high frequency and low amplitude, nevertheless the amplitude diminishes at roughly 0.2 mV, while the period of the high-frequency oscillation is 0.4 mV, which is consistent with Coulomb oscillations.}
    \label{fig:nu8}
\end{figure*}

In Fig.~\ref{fig:nu8} we show conductance oscillations in sample D5. This sample has an estimated twist angle of 0.65$^{\circ}$, where the insulating features appear at $\nu = \pm 8$ \cite{cao2018correlated}. On both electron and hole side, we find an oscillating component with a low amplitude (note that the colorbar scale in Fig.~\ref{fig:nu8}(b) is lower than for the other power spectra presented in this work). It is clear that the frequency of these components increase as a function of gate voltage, and that their frequency increases when the Fermi level approaches $\nu = 8$. This may indicate that spatially correlated charge carriers are important in this system as well. We could, however, not obtain a good fit from Eq.~(8). Possibly, this could indicate that the area $A$ of the confinement is changing, or the interaction energy diminishes faster with increasing carrier density due to the smaller Wigner-Seitz radius. 

\clearpage
\section*{S6: Coulomb oscillations in a relaxed sample}\label{sec:controlsample}
\begin{figure*}[h!]
    \centering
    \includegraphics{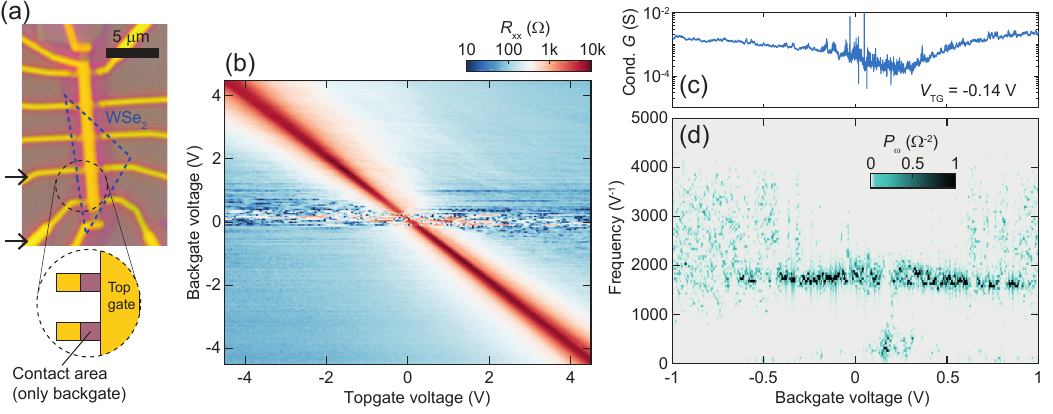}
    \caption{Coulomb oscillations in a sample without moir\'e physics. (a) Optical image of device D6. The blue dotted line indicates the region where WSe$_2$ was incorporated into the stack. (b) 4-point resistance between the two contacts highlighted in panel (a) as a function of backgate and topgate voltage. (c) Two-point conductance as a function of backgate voltage at a fixed topgate voltage. (d) Power spectrum as a function of backgate voltage near the charge neutrality point. }
    \label{fig:controlsample}
\end{figure*}

Figure~\ref{fig:controlsample}(a) shows an optical image of device D6. During fabrication, this device was twisted at 1.3$^{\circ}$, but the device relaxed back to Bernal stacking, which is evidenced by the absence of satellite peaks in Fig.~\ref{fig:controlsample}(b). Furthermore, a top-gate was integrated in the device. An additional WSe$_2$ was incorporated into the stack, which breaks the inversion symmetry of the bilayer graphene and opens a bandgap. Despite the absence of moir\'e physics, we observe high-frequency Coulomb oscillations near the charge neutrality point [Figs.~\ref{fig:controlsample}(c)-\ref{fig:controlsample}(d)]. Since the oscillations remain unchanged as a function of top-gate voltage, they likely emerge from a contact region that is only affected by the backgate [see Fig~\ref{fig:controlsample}(a)]. Since the frequency, and therefore the charging energy, is similar in value to the frequency components in the main part of this work, this sample can serve as a control experiment. From Fig.~\ref{fig:controlsample}(d), it is clear that the frequency remains constant as a function of gate voltage. This is in stark contrast to our observation on devices which show moir\'e satellite peaks, where the frequency increases at low carrier densities. This experiment therefore demonstrates that the band flattening in tBLG is crucial to observe a negative capacitance contribution.

\clearpage
\section*{S7: Observation of Coulomb oscillations in sample D4}\label{sec:rings}
\begin{figure*}[ht!]
    \centering
    \includegraphics{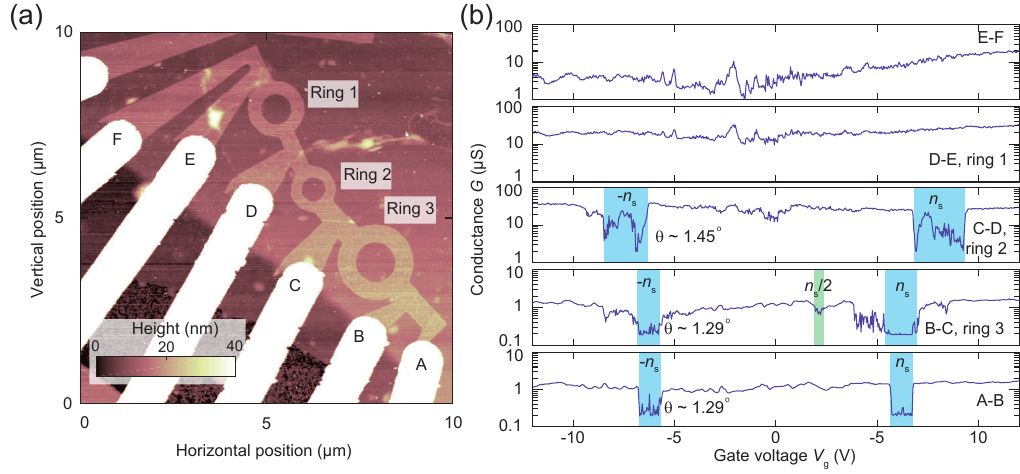}
    \caption{2-point conductance on sample D4. (a) Atomic force microscope image of sample D4, showing the tBLG sample in ring-shaped structures. (b) 2-point conductance measured between different contact pairs and the estimated twist angle.}
    \label{fig:sample4_sample}
\end{figure*}

In Fig.~\ref{fig:sample4_sample}(a) we show an atomic force microscopy map of sample D4, which consists of ring structures. Sample D4 showed significantly higher contact resistances than the other sample due to a different fabrication approach.
Nevertheless, insulating states are resolved as shown in the 2-point conductance traces in Fig.~\ref{fig:sample4_sample}(b). We observe insulating states that point to tBLG in three regions of the device. We estimate the twist angle from the hBN thickness ($35.0$~nm) and a parallel plate capacitor model, since the large contact resistances prevented us from making a more accurate estimation from magnetoresistance measurements. 

\begin{figure*}[ht!]
    \centering
    \includegraphics{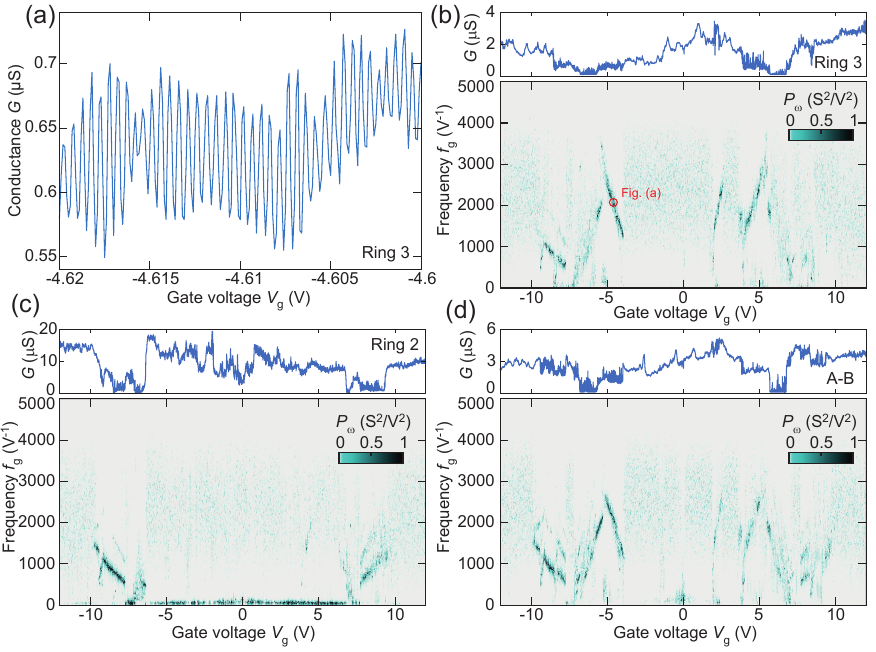}
    \caption{Conductance oscillations observed in sample D4. (a) Conductance oscillations observed in an insulating state on Ring 3. (b) Power spectrum and conductance over the whole gate voltage range on Ring 3, (c) on Ring 2 and (d) on the section between contacts A-B.}
    \label{fig:sample4_oscillations}
\end{figure*}
In Fig. \ref{fig:sample4_oscillations} we show the conductance oscillations observed in the 3 regions with insulating states. This sample also features Coulomb oscillations with perfect periodicity, suggesting that the charging energy also overcomes the quantum level spacing. 

\begin{figure*}[ht!]
    \centering
    \includegraphics{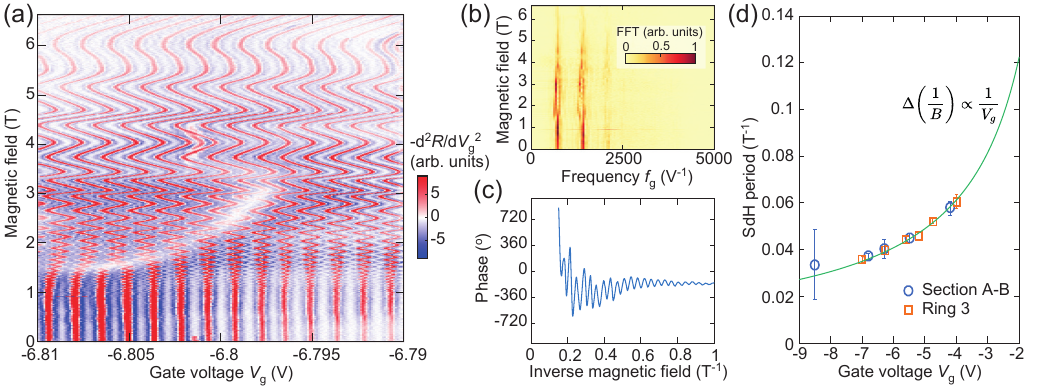}
    \caption{Magnetic field dependence of the oscillations observed in sample D4. (a) Conductance oscillations as a function of magnetic field and gate voltage. (b) Power spectrum at each magnetic field, showing that the frequency of the oscillation remains unchanged. (c) Phase extracted from the power spectrum, showing that the oscillations are periodic in $1/B$. (d) Period of the oscillations as a function of gate voltage, showing that $\Delta\left( \frac{1}{B} \right) \propto \frac{1}{V_g}$.}
    \label{fig:sample4_phase}
\end{figure*}

In sample D4 we were able to reproduce the magnetic field dependence of the oscillations, as outlined in the main text. In Fig.~\ref{fig:sample4_phase} we show that the phase of the oscillations tunes in an identical manner due to the quantum oscillation of the density of states in the region surrounding the confined region. In this case, we also find that the frequency is proportional to the gate voltage, which was also observed on constriction C1.

\clearpage
\section*{S8: Coulomb oscillations in reversed magnetic field}
\begin{figure*}[h!]
    \includegraphics[]{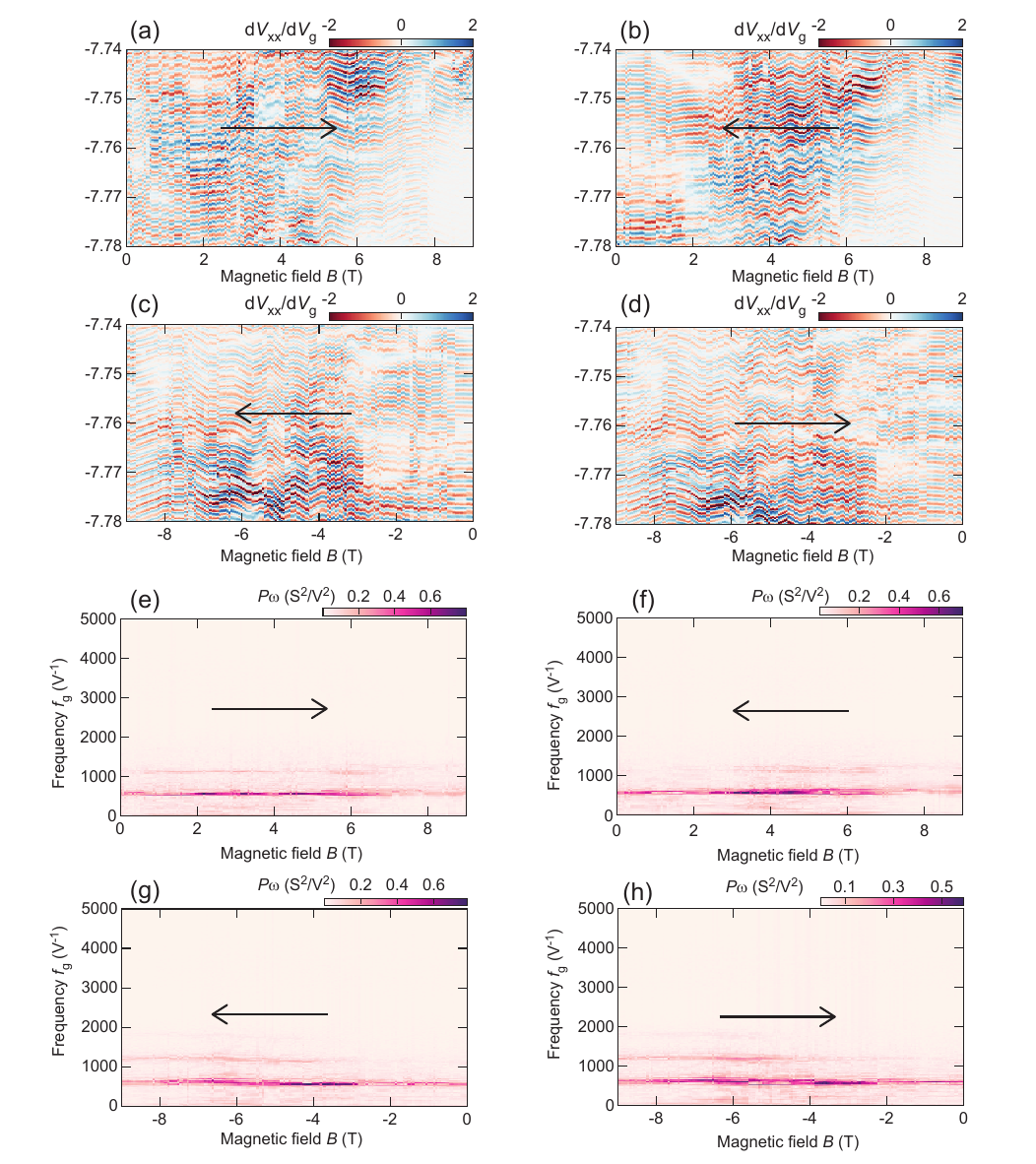}
    \caption{Coulomb oscillations as a function of magnetic field in sample D4. (a--d) Gradient of the voltage drop as a function of magnetic field, sweeping from: (a) 0 T to 9 T, (b) 9 T to 0 T, (c) 0 to -9 T and (d) -9 T to 0 T. (e--h) Frequency content of the oscillations, sweeping the magnetic field from: (e) 0 T to 9 T, (f) 9 T to 0 T, (g) 0 to -9 T and (h) -9 T to 0 T. \label{fig:reversefield}}
\end{figure*}

Figs.~\ref{fig:reversefield}(a)--\ref{fig:reversefield}(d) present Coulomb blockade oscillations obtained from device D4, specifically from the section labeled "Ring 3". 
In this instance, the twist angle deviates somewhat from the magic angle, estimated at 1.29$^{\circ}$, and the frequency components do not align well with Eq.~\ref{eq:slopeequation}. 
Nonetheless, the noticeable increase in frequency as we approach the insulating states hints at a considerable negative contribution to capacitance even in this scenario (Fig.~\ref{fig:sample4_oscillations}).
Examining the frequency content of the signal in Figs.~\ref{fig:reversefield}(e)--\ref{fig:reversefield}(f), we observe no significant alteration in oscillation frequency concerning changes in the magnetic field or when reversing the magnetic field direction.

\clearpage
\section*{S9: Magnetic field dependence of the negative capacitance contribution}
\begin{figure*}[h!]
    \centering
    \includegraphics{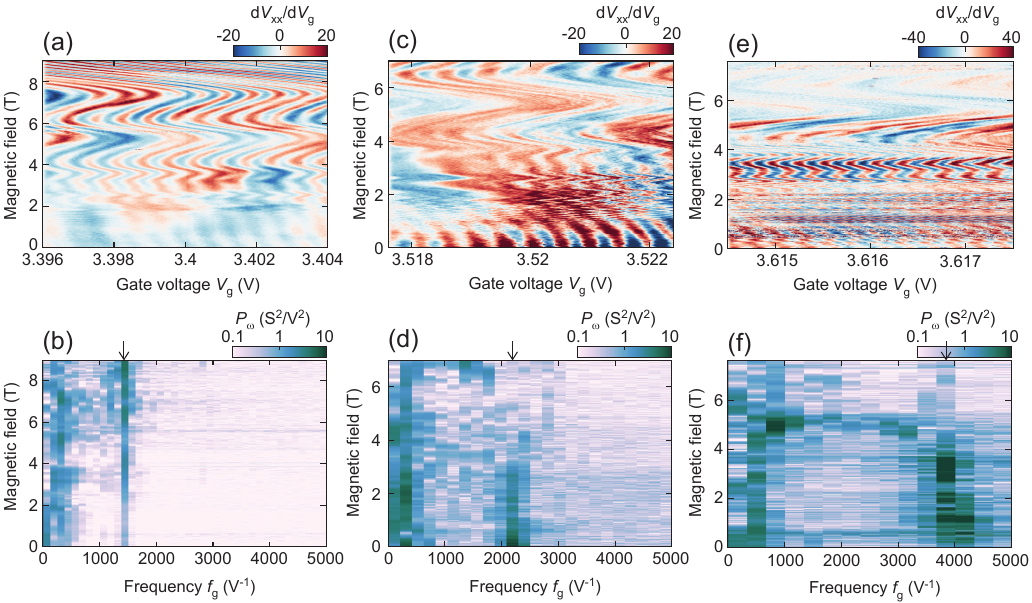}
    \caption{ (a,c,e) Coulomb oscillations as a function of magnetic field near $V_g = 3.4$~V, $V_g = 3.52$~V and $V_g = 3.616$~V, respectively. (b,d,f) Fast-Fourier-transform of the oscillations as a function of magnetic field near $V_g = 3.4$~V, $V_g = 3.52$~V and $V_g = 3.616$~V, respectively. }
    \label{fig:magfield}
\end{figure*}

In this section we present a more detailed investigation of the Coulomb oscillations associated with the frequency component C1.3 under the influence of an out-of-plane magnetic field.
In Fig.~\ref{fig:magfield}(a), \ref{fig:magfield}(c), and \ref{fig:magfield}(e), we sweep the gate voltage within a narrow range around specific values [$V_g = 3.4$ V, $3.52$ V, and $3.161$ V as indicated in Fig.~\ref{fig:observe_oscillations}(f)] while incrementally increasing the magnetic field. 
Once again, we clearly observe quantum oscillations, evident as shifts in peak positions.
Furthermore, the separation between the peaks, corresponding to the frequency of the Coulomb oscillations, remains entirely consistent despite the increase in magnetic field strength. 
This stability is further affirmed by the power spectra depicted in Figs.~\ref{fig:magfield}(b), \ref{fig:magfield}(d), and \ref{fig:magfield}(f), where the frequency component C1.3 is clearly identifiable in each spectrum (indicated by an arrow).
It is important to note that while the amplitude of the oscillations may vary, we do not observe a continuous shift in the frequency, which would suggest a change in the polarization of the electron gas. 
Instead, frequency components appear or disappear as a function of the magnetic field. 
This phenomenon can be attributed to multiple charge island being present on the samples, wherein the prominence of the Coulomb blockade changes with the magnetic field.
However, in Fig.~\ref{fig:magfield}(b) and \ref{fig:magfield}(f), the component C1.3 appears both at 0~T and 9~T, demonstrating that the frequency of an individual charge island remains constant.

\end{document}